\RequirePackage{amsmath}
\documentclass[a4paper,fleqn,usenatbib]{mnras}

\usepackage{mathptmx}
\usepackage{txfonts}

\usepackage[T1]{fontenc}
\usepackage{ae,aecompl}

\usepackage{graphicx}	% Including figure files
\usepackage{amsmath}	% Advanced maths commands
\usepackage{amssymb}	% Extra maths symbols
\usepackage{subfig}
\usepackage{bm}
\usepackage{rotate}
\usepackage{rotating}
\usepackage{verbatim}
\newif\ifAMStwofonts
\AMStwofontstrue

\def\apj{\rmfamily{ApJ~}}         % Astrophysical Journal
\def\apjl{\rmfamily{ApJ~}}        % Astrophysical Journal, Letters
\def\apjs{\rmfamily{ApJS~}}       % Astrophysical Journal, Supplement

\def\aap{\rmfamily{A\&A~}}        % Astronomy and Astrophysics

\def\mnras{\rmfamily{MNRAS~}}     % Monthly Notices of the Royal Astron. Soc.
\def\nat{\rmfamily{Nature~}}      % Nature
\def\ao{\rmfamily{Appl.~Opt.~}}   % Applied Optics

\def\procspie{\rmfamily{Proc.~SPIE~}}% Proceedings of the SPIE
\def\gs{\mathrel{\hbox{\rlap{\hbox{\lower4pt\hbox{$\sim$}}}\hbox{$>$}}}}
\def\ls{\mathrel{\hbox{\rlap{\hbox{\lower4pt\hbox{$\sim$}}}\hbox{$<$}}}}

\def\xmm{{\it XMM-Newton}}

\def\mrk335{{Mrk~335}}

\def\ga{{\rm\thinspace gauss}}

\title[ The geometry \& motion of AGN coronae ]{ Probing the geometry and motion of AGN coronae through accretion disc emissivity profiles }

\author[A. G. Gonzalez et al.]{
A. G. Gonzalez$^{1}$\thanks{E-mail: agonzalez@ap.smu.ca},
D. R. Wilkins$^{1,2}$,
L. C. Gallo$^{1}$
\\
$^{1}$Department of Astronomy and Physics, Saint Mary's University, 923 Robie Street, Halifax, NS, B3H 3C3, Canada\\
$^{2}$Kavli Institute for Particle Astrophysics and Cosmology, Stanford University, 452 Lomita Mall, Stanford, CA 94305, USA
}

\date{Accepted 2017 August 9. Received 2017 August 9; in original form 2016 August 24}

\pubyear{2017}

\begin{document}
\label{firstpage}
\pagerange{\pageref{firstpage}--\pageref{lastpage}}
\maketitle

\begin{abstract}
To gain a better understanding of the inner disc region that comprises active galactic nuclei it is necessary to understand the pattern in which the disc is illuminated (the emissivity profile) by X-rays emitted from the continuum source above the black hole (corona). The differences in the emissivity profiles produced by various corona geometries are explored via general relativistic ray tracing simulations. Through the analysis of various parameters of the geometries simulated it is found that emissivity profiles produced by point source and extended geometries such as cylindrical slabs and spheroidal coronae placed on the accretion disc are distinguishable. Profiles produced by point source and conical geometries are not significantly different, requiring an analysis of reflection fraction to differentiate the two geometries. Beamed point and beamed conical sources are also simulated in an effort to model jet-like coronae, though the differences here are most evident in the reflection fraction. For a point source we determine an approximation for the measured reflection fraction with the source height and velocity. Simulating spectra from the emissivity profiles produced by the various geometries produce distinguishable differences. Overall spectral differences between the geometries do not exceed 15 per cent in the most extreme cases. It is found that emissivity profiles can be useful in distinguishing point source and extended geometries given high quality spectral data of extreme, bright sources over long exposure times. In combination with reflection fraction, timing, and spectral analysis we may use emissivity profiles to discern the geometry of the X-ray source.
\end{abstract}


%%%%%%%%%%%%%%%%%%%%%%%%%%%%%%%%%%%%%%%%%%%%%%%%%%

%%%%%%%%%%%%%%%%% BODY OF PAPER %%%%%%%%%%%%%%%%%%

%######################%
\section{Introduction} %
%######################%
\label{sect:intro}

Active galactic nuclei (AGN) are responsible for some of the most energetic and luminous X-ray phenomena in the Universe. The spectra observed from these objects are produced by photons emitted by the hard X-ray source that lies above the optically thick, geometrically thin accretion disc. Thermal seed photons emitted by the accretion disc are inverse-Compton scattered in the hot electron plasma (corona) above the disc and are emitted as X-rays forming a power law continuum \citep{Sunyaev1979}, which may in turn be reflected off of the accretion disc producing an observed reflection spectrum \citep{George1991}. 

The formation of the corona is thought to be a result of the acceleration and confinement of energetic particles by magnetic fields anchored on the accretion disc \citep{Galeev1979, Haardt1991, Merloni2001}. Its geometry is likely to be complex (e.g. patchy or collimated) (e.g. \citealt{Wilkins2015b, Wilkinsetal2015}), extended over some part of the inner disc (e.g. \citealt{WF12}; henceforth WF12), and dynamic on various time scales (e.g. \citealt{Gallo2015, Wilkins2015a}). However, for modelling purposes it is often considered a point-like source hovering over the black hole spin axis (i.e. the ``lamppost'' model). This is not an unreasonable simplification since current spectral (e.g. \citealt{Brenneman2006, Gallo2013, Parker2015}) and timing studies (e.g. \citealt{Fabian2009, Fabian2013, Zoghbi2010}) do support a compact and centrally located corona, though in a growing number of cases this is insufficient (e.g. WF12, \citealt{Wilkins2015a}).

By studying the illumination pattern of the accretion disc by the corona (emissivity profile) it is possible to probe the nature of the source and its geometry. Throughout this work the illumination profile (photons incident on the disc) and emissivity profile (photons processed and re-emitted by the disc) are treated as equivalent, and thus the terms used interchangeably. \cite{Laor1991} defined the emissivity, $\epsilon\left(r\right)$, as the reflected flux from the accretion disc as a function of radius measured in the frame of the accretion disc. The reflection spectrum observed at infinity, $F_0\left(\nu_0\right)$, is therefore of the form:
\begin{align}
F_0\left(\nu_0\right) & = \int{\epsilon\left(r_e\right)I_r\left(\frac{\nu_0}{g}\right)T\left(r_e,g\right)dg\,r_edr_e}
\end{align}
where $I_r$ is the rest-frame reflection from the disc, $T\left(r_e,g\right)$ is the transfer function which projects the rays to the observer, and is integrated over all redshift values $\mathrm{g = E_{\mathrm{em}} / E_{\mathrm{obs}}}$, where $E_{\mathrm{em}}$ is the energy emitted and $E_{\mathrm{obs}}$ is the energy measured by an observer, over all radii $r$ of the accretion disc such that the emitted line is shifted to $\nu_{0}$. The emissivity profile can be measured from AGN data and fit with a power law in order to gain insight into the nature of the X-ray source (e.g. \citealt{Wilkins2011}). Simulating various corona geometries and exploring the emissivity profiles produced allows for comparisons to be made between theoretical models and the measured profiles obtained from AGN data.

Modelling the emissivity profiles produced by point source coronae (e.g. \citealt{Miniutti2003}), ring-like geometries (e.g. \citealt{Suebsuwong2006}), and extended slabs (which may be seen as representing the bulk of a more detailed extended model) (e.g. WF12) has been shown to accurately reproduce these measured profiles. WF12 were able to successfully show through comparing simulated emissivity profiles with observational data that the extended nature of the corona can be detected in the emissivity profile of the blurred reflection spectrum.

For the narrow line Seyfert 1 (NLS1) galaxy 1H~0707--495 the emissivity profile was consistent with a twice-broken power law. The profile was steep in the inner region where light bending is dominant, flat at intermediate radii where the corona blankets the disc, and drops with $r^{-3}$ at large distances. Further analysis shows the emissivity profile, and hence the geometry of the corona, may change depending on the flux of the source.  For example, \cite{Wilkins2014} found the corona in 1H~0707--495 to be extended in the bright state but compact when the source was dim, showing a dynamic nature of the corona undergoing phase changes and altering its geometry significantly during these processes. The emissivity profile is a strong probe for geometry of the corona that can be used alongside various other tools including the reflection fraction (e.g. \citealt{WilkinsANProc2015,Wilkins2015a}) and time-lag analysis (e.g. \citealt{Wilkins2013}) in order to gain further insights into the nature of the X-ray source.

The goal of this work is to extend upon the growing research base on extended corona geometries by simulating various sources and examining the differences between them. In this paper we will review the point source corona before moving into extended corona geometries, including a section on beamed point and extended sources. We investigate various parameters of the corona geometry to determine which are visible in the produced emissivity profiles. We also examine other spectral signatures of the extended coronae that may be measured in observations.

%#####################################%
\section{Ray tracing \& Calculations} %
%#####################################% 
\label{sect:raytracing}

The theoretical emissivity profiles shown throughout this paper were produced via general relativistic ray tracing simulations of various corona geometries above a rotating Kerr black hole. The Kerr metric in Boyer-Lindquist co-ordinates follows as:
\begin{align}
\label{eqn:kerr}
\begin{split}
ds^2 &= c^2\left(1 - \frac{2\mu r}{\rho^2} \right)dt^2 + \frac{4\mu a c r \sin^2 \theta}{\rho^2}dtd\varphi - \frac{\rho^2}{\Delta}dr^2 -\rho^2d\theta^2 \\
&-\left(r^2 + a^2 + \frac{2\mu a^2 r \sin^2\theta}{\rho^2}\right) \sin^2\theta d\varphi^2
\end{split}
\end{align}
where $a \equiv J/Mc$, $\mu \equiv GM/c^2$, $\rho^2 \equiv r^2+a^2\cos^2\theta$, and $\Delta \equiv r^2-2\mu r +a^2$. Making use of the convention $c = \mu = 1$ provides units of gravitational radii such that $r_g = GM/c^2$.

The exact method in which the ray tracing simulations were performed, along with the construction of the X-ray source, is presented in WF12 and is summarized briefly here. 

Photons are propagated along null geodesics in the spacetime surrounding a Kerr black hole until they reach the accretion disc, the event horizon, or escape the system. The X-ray source is constructed based on the geometry to be studied (e.g. a point source, cylindrical, spheroidal, ellipsoidal, or conical geometry) with rays starting a random locations within the defined source region. A tetrad of basis vectors, with one time-like ($\mathbf{e}'_{(t)}$) and three space-like ($\mathbf{e}'_{(1)}$, $\mathbf{e}'_{(2)}$, $\mathbf{e}'_{(3)}$) vectors, is constructed for each source type (e.g. stationary, rotating, or beamed) ensuring that they satisfy:
\begin{align}
\mathbf{e}'_{(\alpha)}\cdot\mathbf{e}'_{(\beta)} &= g_{\mu \nu}{e}^{\mu}_{(\alpha)}{e}^{\nu}_{(\beta)} = \eta_{(\alpha)(\beta)}
\end{align}
where $g_{\mu \nu}$ is the metric tensor corresponding to the Kerr metric in Equation \ref{eqn:kerr} and $\eta_{(\alpha)(\beta)}$ is the Minkowski metric. 

Once the rays have been initiated within the source region and propagated to the accretion disc (those that reach the event horizon or escape the system are disregarded) they are summed in radial bins for annuli of area $A\left(r,dr\right)$ on the disc, giving the number of photons as $N\left(r,dr\right)$. Accounting for redshift effects on photon energy dividing by a factor of $g^2$ gives the emissivity profile as:
\begin{align}
\label{eqn:profile}
\epsilon\left(r\right) &= \frac{N\left(r,dr\right)}{g^2 A(r,dr)}
\end{align}
where $g = E_{\mathrm{em}} / E_{\mathrm{obs}}$ with $E_{\mathrm{em}}$ being the energy emitted and $E_{\mathrm{obs}}$ being the energy measured by an observer. By calculating the emissivity profiles of various corona geometries and analysing the differences between them raises the possibility of detecting different source geometries in AGN data.

%#######################%
\section{Point Sources} %
%#######################%
\label{sect:ptsrc}

\begin{figure}
	\subfloat[\label{fig:ptsrc}]{\includegraphics[width=\linewidth]{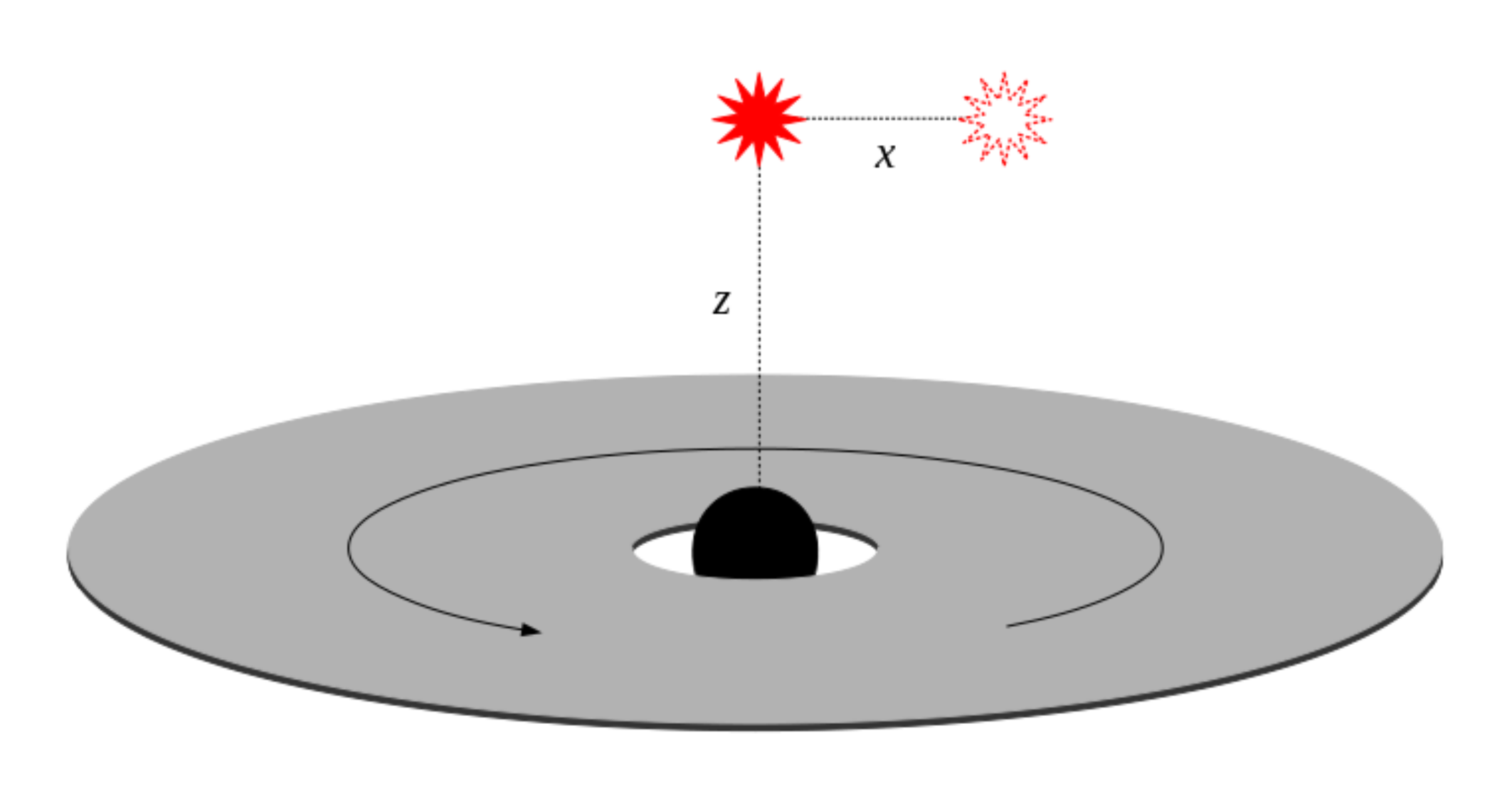}}
	\\
	\subfloat[\label{fig:ptsrc_height_comp}]{\includegraphics[width=\linewidth]{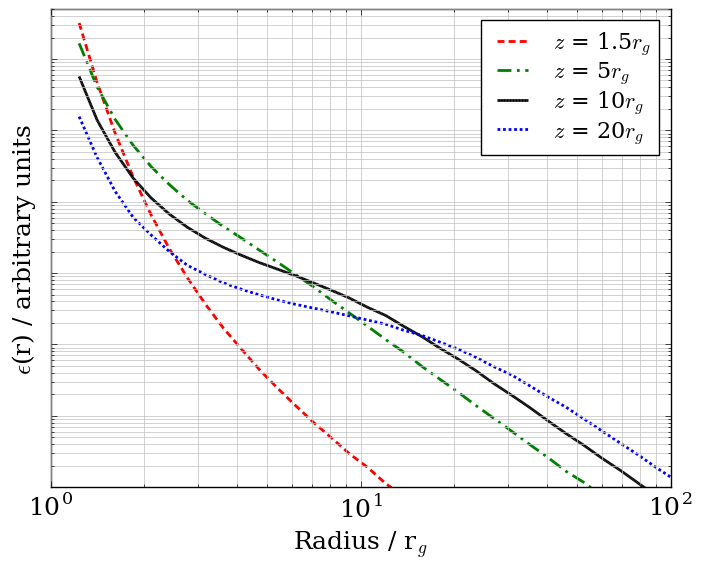}}
	\\
	\subfloat[\label{fig:ptsrc_disp_comp}]{\includegraphics[width=\linewidth]{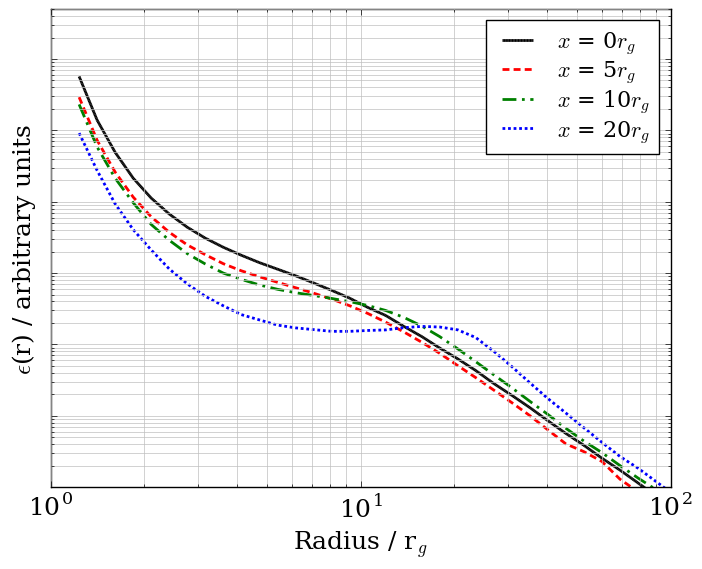}}
	
	\caption{(a) A schematic representation of a stationary, isotropic point source located at a height $z$ with displacement $x$ from the spin axis above a rotating Kerr black hole and accretion disc. (b) A comparison of point sources at four different heights ranging from $z = 1.5r_g$ to $z = 20r_g$ located on the rotation axis of a black hole with $a = 0.998$. (c) Point source coronae at $z = 10r_g$ with $x = 0r_g$ to $x = 20r_g$ away from the rotation axis.}
	\label{fig:ptsrc:all}
\end{figure}

The venture into examining different corona geometries begins with the simple stationary, isotropic point source (the aforementioned ``lamppost'' model). In the simulations, the source is held at a fixed height $z$ on the spin axis above the accretion disc of a rotating Kerr black hole with dimensionless spin parameter $a = 0.998$, as shown in Figure \ref{fig:ptsrc}.

As in WF12 a set of basis vectors must be constructed for this source model in order to evaluate the energies of the emitted photons as well as the angle of photon emission according to the rest frame of an observer outside of the source frame of the corona. In this framework, the space-like basis vectors are such that they form a Cartesian coordinate system in the observer's frame. The source may also be given a rotational velocity $\Omega = d\varphi/dt = \left(a \pm x^{3/2}\right)^{-1}$ about the spin axis, where $x$ is the distance from the spin axis. For a stationary, isotropic point source $\Omega = 0$. Constructing the tetrad of basis vectors, using the method in Appendix A of WF12, for a point source yields:

\begin{align}
\begin{split}
\label{eqn:rot-basis}
\mathbf{e}'_{(t)} &= \sqrt{\frac{1}{g_{tt}+2\Omega g_{\varphi t}+\Omega ^2 g_{\varphi\varphi}}} \left( 1, 0, 0, \Omega \right) \\
\mathbf{e}'_{(1)} &= \left( 0, \sqrt{\frac{-1}{g_{rr}}}, 0, 0 \right) \\
\mathbf{e}'_{(2)} &= \left( 0, 0, \sqrt{\frac{-1}{g_{\theta\theta}}}, 0 \right) \\
\mathbf{e}'_{(3)} &= \sqrt{\frac{1}{g_{tt}+2\Omega g_{\varphi t}+\Omega ^2 g_{\varphi\varphi}}} \\ 
& \quad\quad\quad\quad\quad \left( g_{\varphi t} + \Omega g_{\varphi\varphi}, 0, 0, \Omega g_{\varphi t} - g_{tt}\right)
\end{split}
\end{align}

Ray tracing simulations were run for stationary, isotropic point sources at $z = 1.5r_g$, $z = 5r_g$, $z = 10r_g$, and $z = 20r_g$. The results of these simulations are shown in Figure \ref{fig:ptsrc_height_comp}. 

The general shape of all lines on the plot resembles a twice-broken power law with a steeply falling inner profile and an outer profile following $r^{-3}$. The steep slope of the inner profile is a result of time dilation greatly enhancing the flux of photons incident on the innermost region of the accretion disc according to observers closest to the black hole. The profile shape for point sources obtained here as well as the trend with increased height agree with similar works studying the lamppost geometry (e.g. \citealt{Miniutti2003,Suebsuwong2006};WF12;\citealt{Dauser2013,Dovciak2014}).

Trials with spin parameter $a < 0.998$ act to increase the innermost stable circular orbit (ISCO) and therefore truncate the profile at small $r$. The spin parameter is held at $a = 0.998$ for all following work. 

The first property of the point source model is the height of the corona above the black hole. Figure \ref{fig:ptsrc_height_comp} shows how the height of the source effects its emissivity profile. For a point source at $z = 1.5r_g$ the effects of general relativity are extreme. Most of the photons landing on the accretion disc are focused tightly around the black hole and immediate vicinity. Increasing the height of the source produces a twice-broken power law shape in the emissivity profile that becomes pronounced for larger values of $z$, as found in WF12. 

For $r \approx z$ in the latter three cases a turnover exists in the profile at the second break point. This second break-point in the twice-broken power law description is measured directly from the Fe K$\alpha$ line in the reflection spectrum of AGN and can be constrained to within $\sim$15--30 per cent error (e.g. \citealt{Wilkins2011,Wilkins2015b}). Common to all cases is the outer profile which follows $r^{-3}$ as in the case of the classical description of a point source above an infinite plane. 

The second property of the point source geometry is displacement from the spin axis, given by $x$, the results of which are shown in Figure \ref{fig:ptsrc_disp_comp} for a point source located at $z = 10r_g$. As the source moves away from the spin axis the inner disc receives fewer photons while the outer disc receives proportionally more. Particularly, where $r \approx x$ the break in the emissivity profile is accentuated indicating that (other than the inner disc) the largest number of photons land on the disc directly under the point source. If these sources were each to be co-rotating with the accretion disc, a more pronounced break would be visible in the profile as rays are beamed in the direction of motion, focusing the emission and further increasing the number of photons received by the disc at $r \approx x$, as shown in WF12. The axisymmetric nature of the Kerr spacetime means that this example of a displaced point source is equivalent in disc illumination to a ring-like source geometry as we consider the emissivity profiles as a function of $r$ only and are therefore azimuthally averaged.

\begin{figure}
	\scalebox{1.0}{\includegraphics[width=\linewidth]{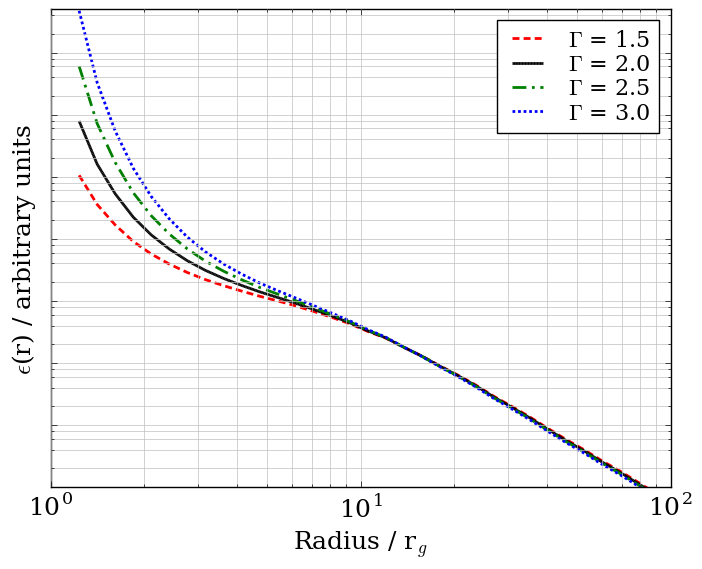}}
	\caption{Emissivity profiles produced by point sources at height $z = 10r_g$ with varying values of photon index $\Gamma$ for the incident spectrum.}
	\label{fig:ptsrc_gamma_comp}
\end{figure}

Thus far, we have been assuming the value of $\Gamma$, the photon index, of the power law spectrum emitted from the corona incident on the accretion disc as $\Gamma = 2.0$. We may vary this value and study its effect on the emissivity profile of a point source, shown in Figure \ref{fig:ptsrc_gamma_comp}. 
The variation of the photon index requires a variant of Equation \ref{eqn:profile} where $g^2$ in the denominator is replaced by $g^{\Gamma}$. This arises as the number of photons travelling along any given ray must be conserved for different $\Gamma$, requiring the energy correction, previously $g^2$, to be adjusted accordingly, to $g^{\Gamma}$, in order to properly account for the change in the number of photons in each bin of the emissivity profile. As can be seen in the plot, by using a spectrum with a steeper continuum we have increased the slope of the profile form the innermost region of the accretion disc. Beyond $r \approx 10r_g$, however, the effect of changing the value of $\Gamma$ quickly becomes minimal. This effect, though important, will not be further studied in the cases to come as the result is well illustrated with this single example. With this preliminary analysis of point sources complete it is now possible to examine more physically plausible examples of extended corona geometries.

%##########################%
\section{Extended Sources} %
%##########################%
\label{sect:extsrc}

Extended source geometries can be simulated as a sum of point sources within a defined region, producing an optically thin extended corona geometry. Each of the rays within the constructed geometry is given a random initial position and random direction with distribution such that each point emits isotropically in its own rest frame. Photons are then propagated along null geodesics, as in Section \ref{sect:ptsrc}, until they either reach the disc, are lost beyond the event horizon, or escape the system. Note that, as in WF12, extended sources are assumed to have a homogeneous emissivity profile. This means that finding the best-fitting profile to an observed spectrum finds the geometry as extent of a structure that is representative of the bulk of the coronal emission.

%########################%
\subsection{Cylindrical} %
%########################%
\label{sect:cylsrc}

\begin{figure*}
	\subfloat[\label{fig:cylsrc}]{\includegraphics[width=0.49\textwidth]{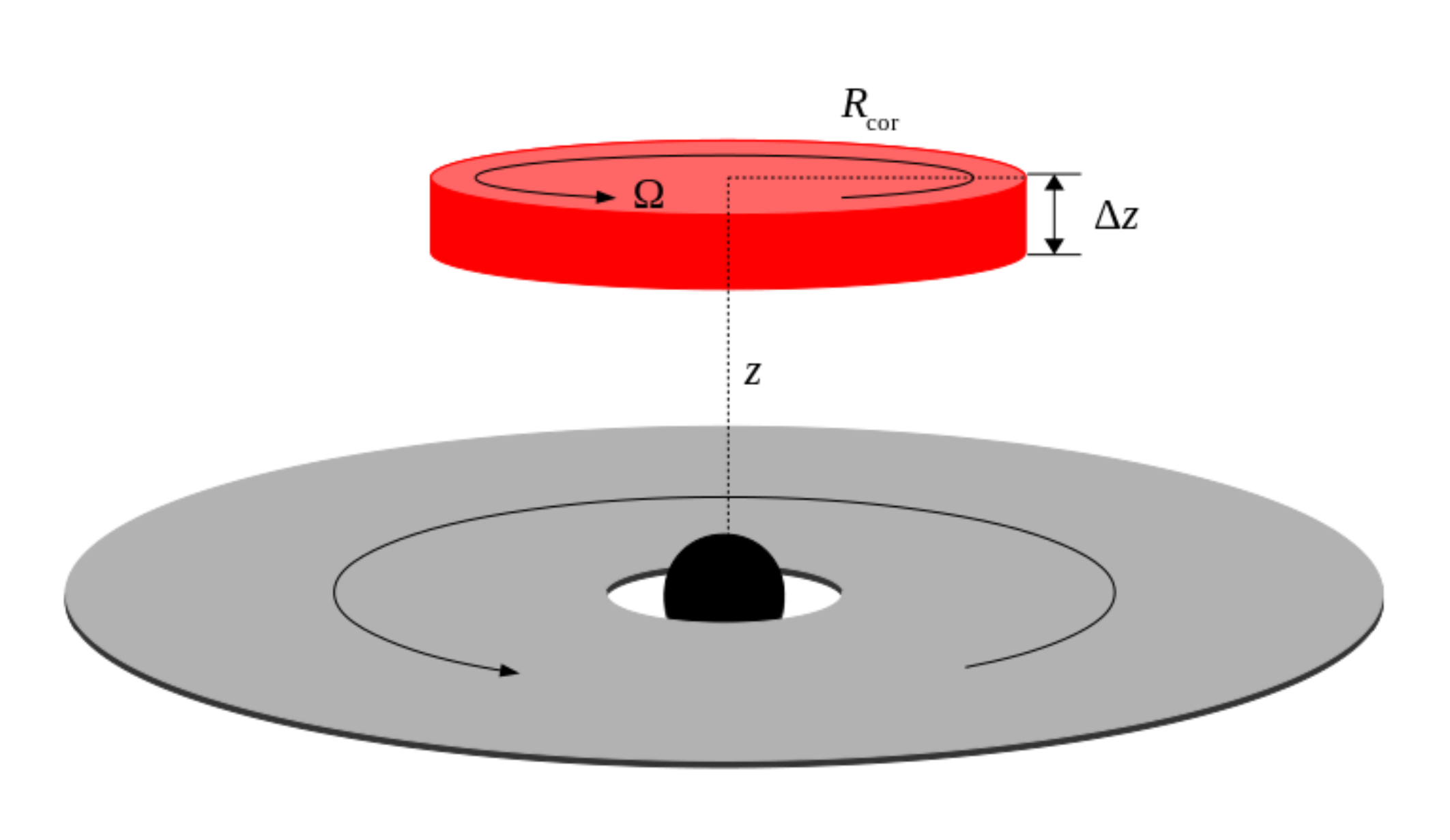}}
	\hfill
	\subfloat[\label{fig:cylsrc_pt_comp_10rg}]{\includegraphics[width=0.49\textwidth]{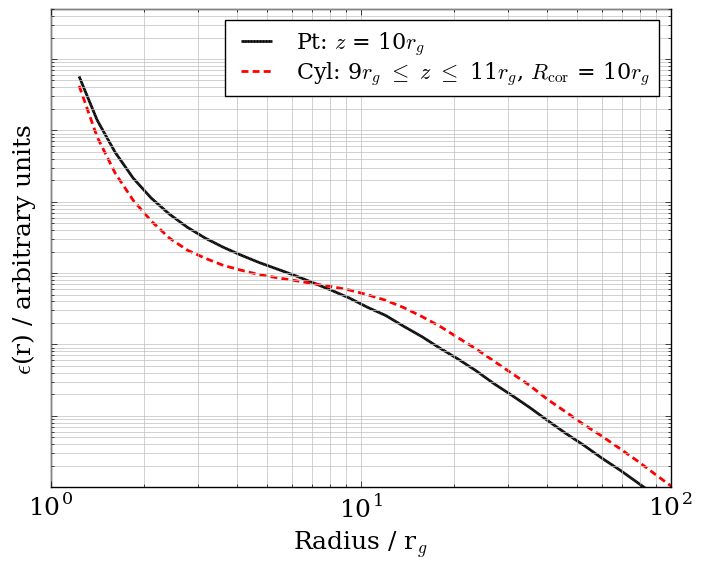}}
	\\
	\subfloat[\label{fig:cylsrc_height_comp}]{\includegraphics[width=0.49\textwidth]{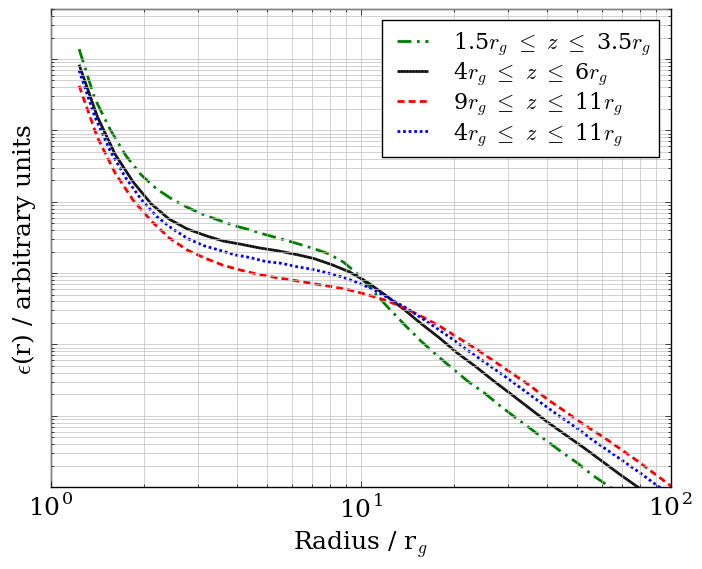}}
	\hfill
	\subfloat[\label{fig:cylsrc_extent_comp}]{\includegraphics[width=0.49\textwidth]{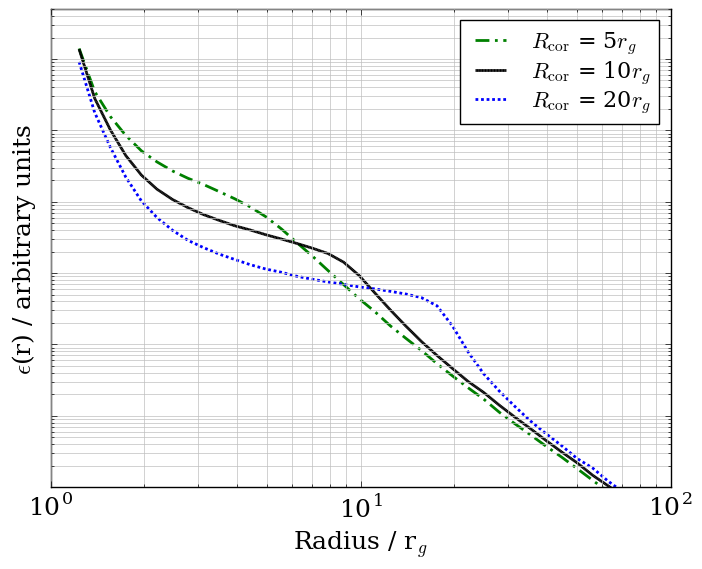}}
	
	\caption{(a) A cylindrical slab corona geometry rotating with $\Omega = d\varphi / dt$ located at a height $z$ above a rotating black hole. It has thickness $\Delta z$ and extends radially a distance $R_{\mathrm{cor}}$ over the accretion disc. (b) A comparison of a point source at $z = 10r_g$ with an extended cylindrical slab centred on $z = 10r_g$ with extent $R_{\mathrm{cor}} = 10r_g$ and thickness $\Delta z = 2r_g$. (c) The effect of height on profiles of cylindrical sources centred on $z = 2.5r_g$, $z = 5r_g$, and $z = 10r_g$ with $R_{\mathrm{cor}} = 10r_g$ and of thickness $\Delta z = 2r_g$, with a thick slab centred on $z = 7.5r_g$ of thickness $\Delta z = 7r_g$. (d) Various cylindrical sources with radial extents $R_{\mathrm{cor}} = 5r_g$, $R_{\mathrm{cor}} = 10r_g$, and $R_{\mathrm{cor}} = 20r_g$ all centred on $z = 2.5r_g$ and of thickness $\Delta z = 2r_g$.}
	\label{fig:cylsrc:all}
\end{figure*}

We begin the extended source geometry analysis with a rotating cylindrical slab at height $z$ above the accretion disc. The cylinder has thickness $\Delta z$ and radial extent $R_{\mathrm{cor}}$, as seen in Figure \ref{fig:cylsrc}. In this geometry points are assumed to co-rotate with the portion of the disc directly beneath, if, for instance, they are accelerated by field lines anchored to the disc beneath (e.g. WF12). This gives points in the corona a rotational velocity of $\Omega = d\varphi/dt = \left(a \pm x^{3/2}\right)^{-1}$ where $x$ is the distance from the spin axis.

In Figure \ref{fig:cylsrc_pt_comp_10rg} we compare the emissivity profile produced by a cylindrical slab, centred on $z = 10r_g$ with radial extent $R_{\mathrm{cor}} = 10r_g$ and thickness $\Delta z = 2r_g$, to a point source at $z = 10r_g$. It can be seen that in the case of the slab corona the photons landing on the accretion disc are spread over the outer radii of the disc causing the profile to be flattened in comparison with that of the point source. This result is to be expected from the extended nature of the slab geometry as we are essentially summing the emissivity profiles of all point sources contained within the region defined by the cylindrical source. 

At just over $r = 10r_g$ a break point in the profile of the disc is produced in the case of the cylindrical corona, corresponding with the outer edge of the slab. The profile then falls off with slope approximately equal to that of the outer profile produced by point source at large radii. The twice-broken power law shape is prevalent in both cases, though the flattened midsection of the emissivity profile resulting from the slab geometry enhances the shape. 

By simulating different source heights above the accretion disc for extended cylindrical models, the emissivity profiles in Figure \ref{fig:cylsrc_height_comp} are produced. Placing the corona closer to the accretion disc results in more of the emission being focused on the inner region of the disc, as seen by the enhanced inner profile for the lowest sources. By moving the slab further away from the disc, to be centred at $z = 10r_g$, the inner region of the accretion disc receives a smaller fraction of the emitted photons. Also included is a thick slab of $\Delta z = 7r_g$ that encapsulates the two other slabs centred on $z = 5r_g$ and $z = 10r_g$. From this we can see that thickness of the corona does not significantly impact the shape of the emissivity profile, retaining its twice-broken shape. Overall, increasing distance from the accretion disc acts to stretch the shape of the profile and thus distribute photons more evenly across the disc.

Extending the cylindrical geometries radially produces the emissivity profiles in Figure \ref{fig:cylsrc_extent_comp}. By reducing the radial extent of the slab the break in the profile moves to a smaller radius, and conversely increasing the extent moves the break point out to larger radii. As noted previously, the break point in the emissivity profile traces the outer edge of the slab which allows us to measure the radial extent of the source, as shown in WF12. 

Comparing the results in Figures \ref{fig:cylsrc_height_comp} and \ref{fig:cylsrc_extent_comp} shows that the extent of the slab produces a larger effect on the emissivity profile than does the height of the source. This indicates that for extended geometries covering a large portion of the disc radial extent is more easily determined than source height and has a more significant impact on the emissivity profile overall, as found in WF12. With an understanding of how changes in location and size of an extended source geometry affect the emissivity profile it is now possible to examine different configurations, such as a hemisphere placed directly on the accretion disc.

%######################%
\subsection{Spheroidal} %
%######################%
\label{sect:sphsrc}

\begin{figure*}
	\subfloat[\label{fig:sphsrc}]{\includegraphics[width=0.49\textwidth]{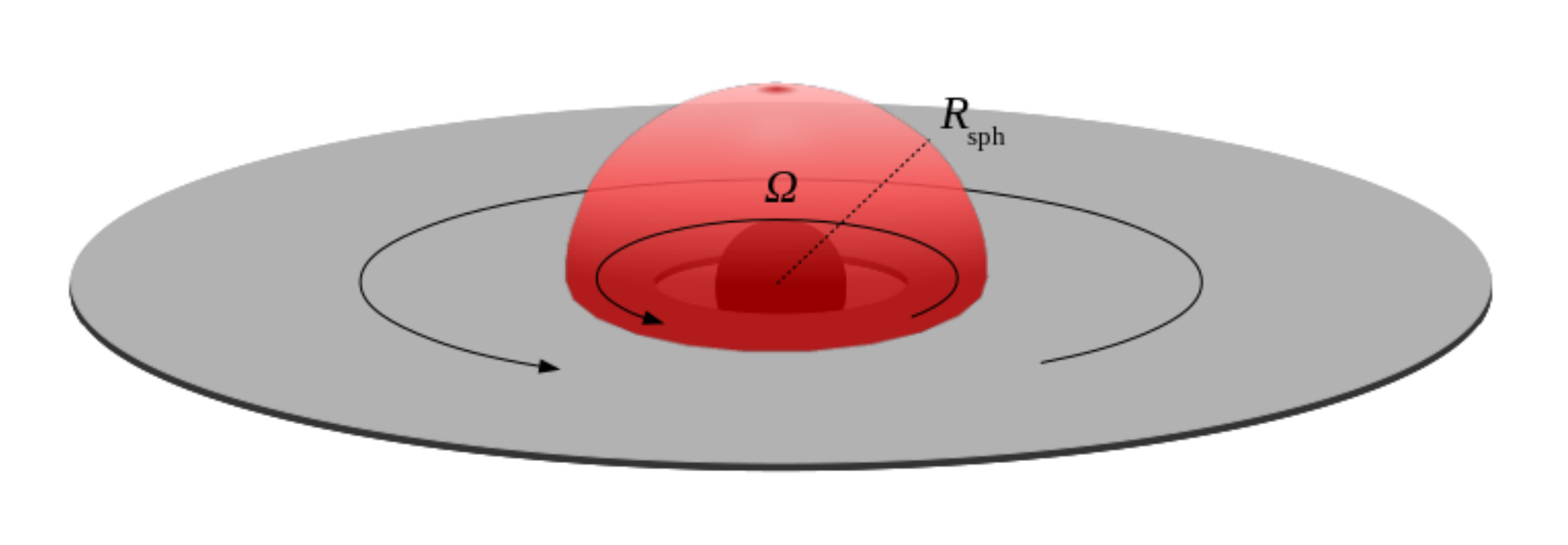}}
	\hfill
	\subfloat[\label{fig:ellsrc}]{\includegraphics[width=0.49\textwidth]{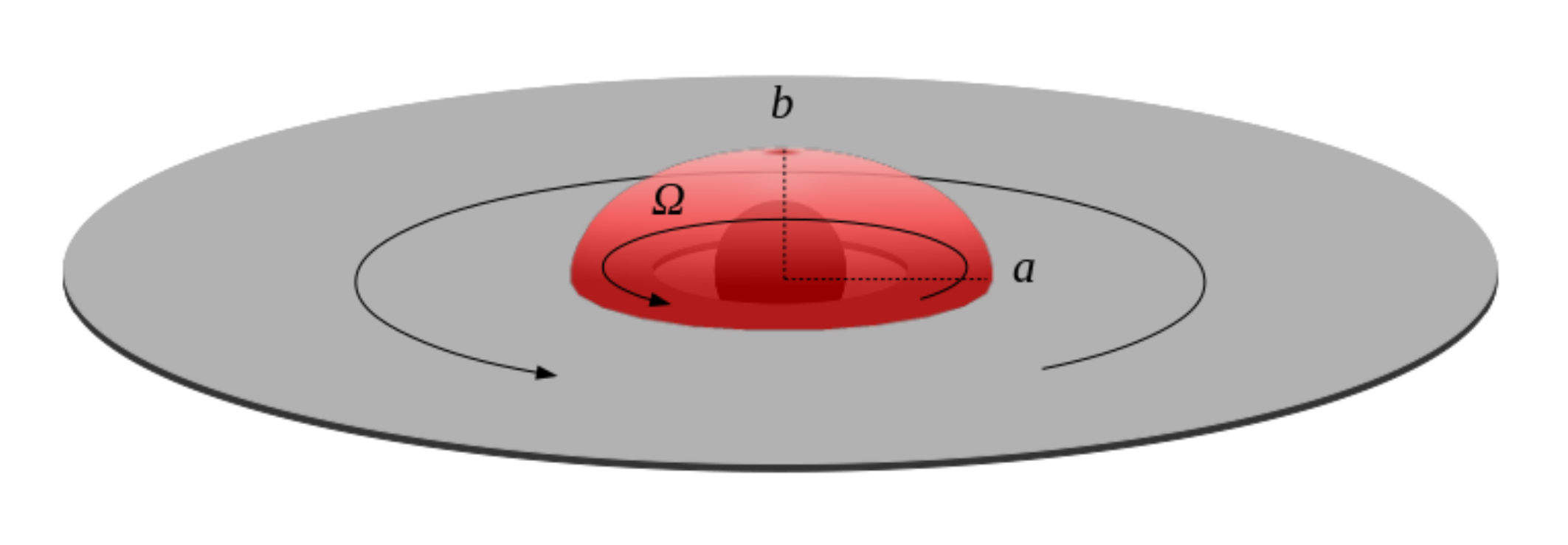}}
	\\
	\subfloat[\label{fig:sphsrc_cyl_comp}]{\includegraphics[width=0.49\textwidth]{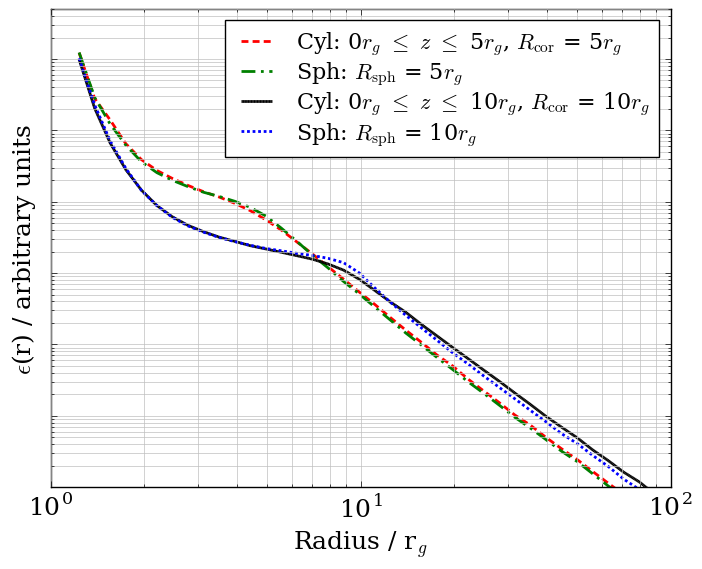}}
	\hfill
	\subfloat[\label{fig:sphsrc_height_comp}]{\includegraphics[width=0.49\textwidth]{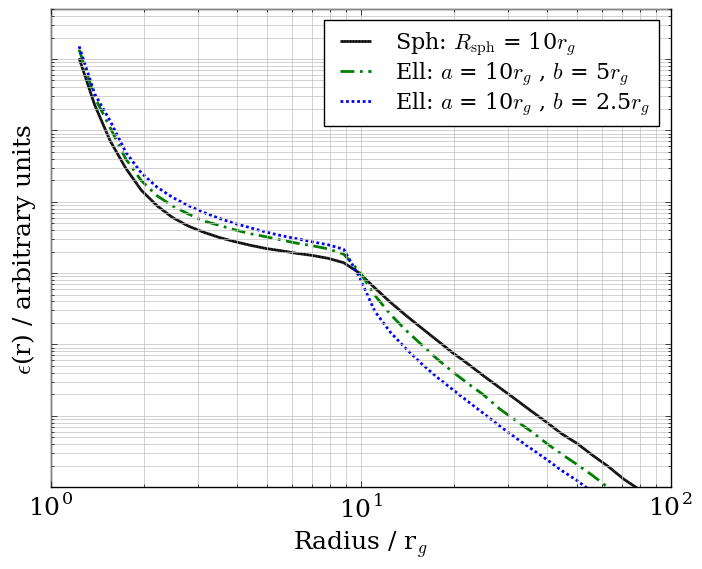}}
	
	\caption{(a) A spheroidal corona geometry rotating with $\Omega = d\varphi / dt$ and radius $R_\mathrm{sph}$ located on the accretion disc of a rotating black hole. (b) An ellipsoid corona geometry rotating with $\Omega = d\varphi / dt$ with semi-major axis $a$ and semi-minor axis $b$ located on the accretion disc of a rotating black hole. (c) Comparing two sets of cylindrical and spheroidal corona geometries of similar size located on the accretion disc. (d) Investigating the effect of height on spheroidal corona geometries with semi-major axes $a = 10r_g$ and varying semi-minor axes $b = 2.5r_g$, $b = 5r_g$, and $b = 10r_g$.}
	\label{fig:sphsrc:all}
\end{figure*}

For the point source and cylindrical slab, the source was located at some height $z$ above the black hole and accretion disc system. In this case a hemisphere of radius $R_{\mathrm{sph}}$ rotating with $\Omega = d\varphi/dt$, as defined in Section \ref{sect:ptsrc}, is placed on the accretion disc, encapsulating the black hole itself, as shown in Figure \ref{fig:sphsrc}. 

Referring back to Figure \ref{fig:cylsrc_height_comp} and the observed effect of moving an extended source closer to the black hole, we already have an intuition that the emissivity profile will exhibit a shortened flatter mid-section as we are close to the black hole. 

Figure \ref{fig:sphsrc_cyl_comp} compares the emissivity profiles produced by two sets of like-sized spheroidal and cylindrical extended geometries placed on the accretion disc. The profiles produced are of similar shape dropping from their flattened midsection at the same radius for like sized sources. Spherical geometries produce a more pronounced break and a decreased number of photons reaching the outer regions of the accretion disc. This result is not unexpected as the spheroidal sources are smaller in volume than similarly sized cylindrical slabs by a factor of one-third, with the missing volume being most significant at the outer edge of the source. As such, the spheroidal geometries emit fewer photons at the maximum radial extents than the cylindrical sources, producing the discussed results in the profiles. 

Moving forward, it is not unreasonable to argue that a spheroidal corona may become oblate due to the orbital motion of the corona. Moreover, if the corona is in fact formed by the magnetic fields anchored on the accretion disc, their movement and rotation would also produce an oblate deformation of the spheroid. Therefore, it is plausible that a spheroidal geometry would warp into an ellipsoidal configuration, with semi-major axis $a$ and semi-minor axis $b$ as in Figure \ref{fig:ellsrc}. 

Comparisons of the emissivity profiles of similarly sized spheroidal and ellipsoidal coronae can be seen in Figure \ref{fig:sphsrc_height_comp}. By decreasing the height $b$ of the spheroidal geometry to become more ellipsoidal, a decreased number of photons reach the outer extent of the accretion disc while an increased number are observed to land on the inner region of the disc. This change produces a more pronounced break point for reductions in the value of $b$. These results are expected as decreasing the height of the spheroidal shape removes a portion of the source located furthest away from the black hole, reducing the number of photons reaching the outer disc. This effect is that fewer photons are able to reach the outer portion of the disc and are emitted closer to the black hole, being drawn into the region of the inner few radii of the accretion disc.

%####################%
\subsection{Conical} %
%####################%
\label{sect:conicalsrc}

In all of the extended geometries studied thus far we have been examining emissivity profiles produced by predominantly radially extended sources. With such corona geometries studied both on the disc and at some height $z$ above it we may move to another physically motivated source configuration with a focus on being vertically extended to simulate the base of a jet. 

\begin{figure*}
	\subfloat[\label{fig:conesrc}]{\includegraphics[width=0.49\textwidth]{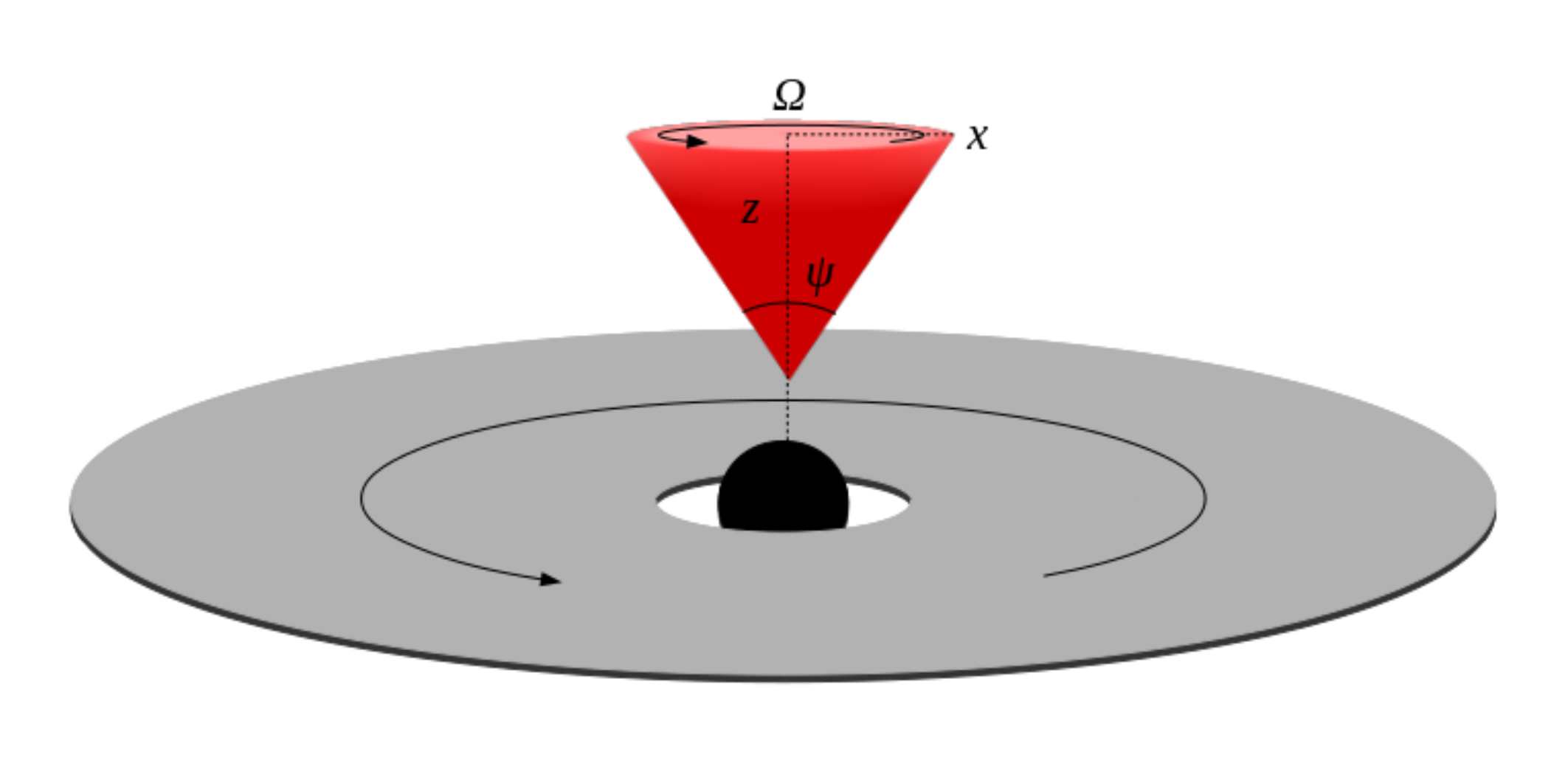}}
	\hfill
	\subfloat[\label{fig:conesrc_pt_comp}]{\includegraphics[width=0.49\textwidth]{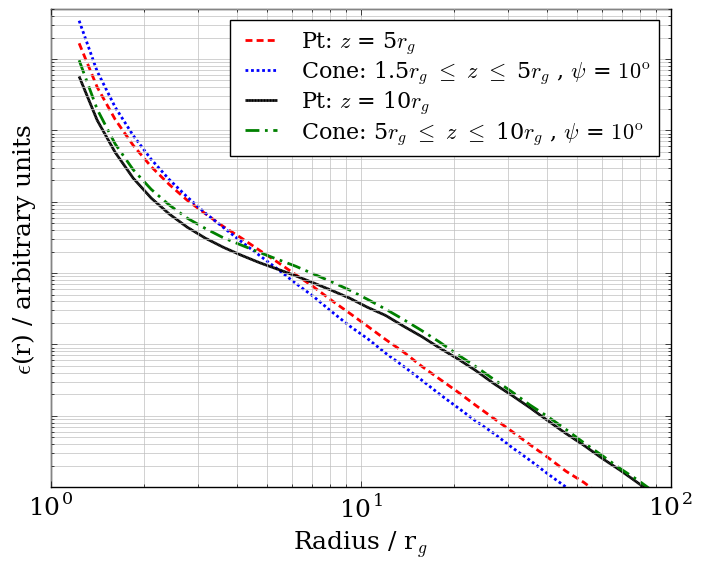}}
	\\
	\subfloat[\label{fig:conesrc_height_comp}]{\includegraphics[width=0.49\textwidth]{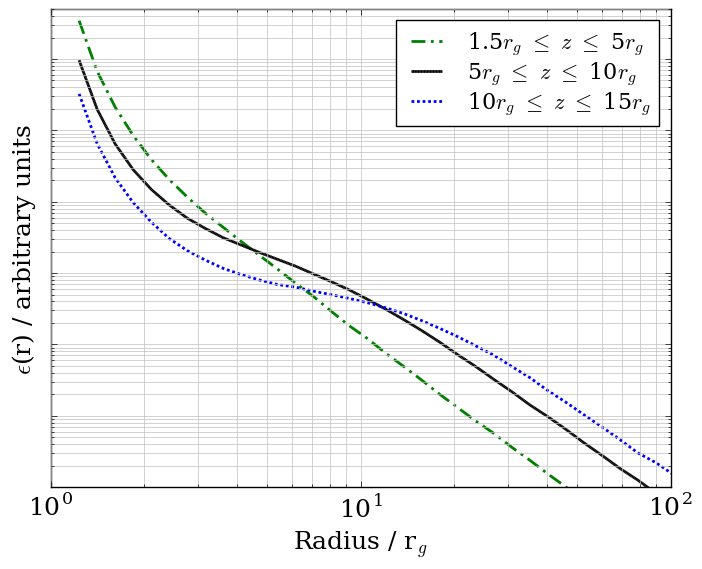}}
	\hfill
	\subfloat[\label{fig:conesrc_angle_comp}]{\includegraphics[width=0.49\textwidth]{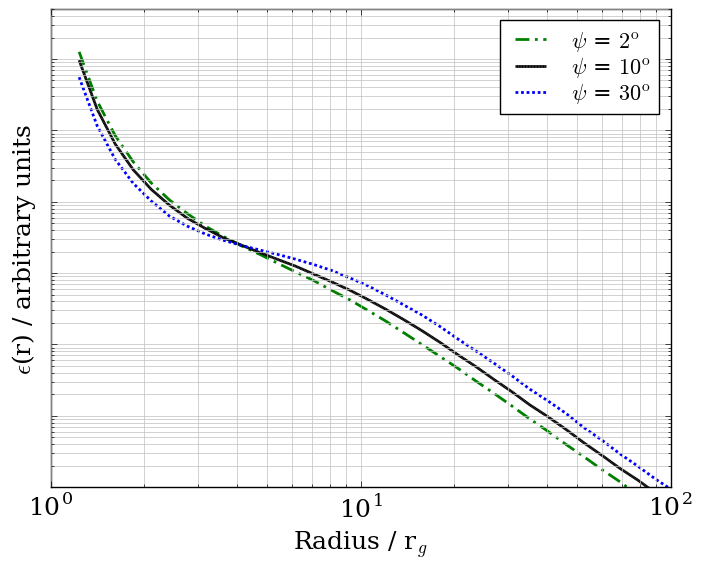}}
	
	\caption{(a) A schematic of the conical corona geometry rotating with $\Omega = d\varphi / dt$ at height $z$ and with opening angle $\psi$ that extends a distance $x$ over the accretion disc. (b) Conical corona geometries compared with point sources placed at the maximum height of the displayed cones with opening angles $\psi = 10^{\circ}$ located a distance $z$ above the black hole. (c) The effect of source height on the emissivity profiles of conical sources with $\psi = 10^{\circ}$. (d) The effect of opening angle on the shape of the emissivity profile for conical source with $\psi = 2^{\circ} - 30^{\circ}$ and common heights $5r_g \leq z \leq 10r_g$.}
	\label{fig:conesrc:all}
\end{figure*}

Conical geometries opening away from the black hole pose an attractive and physically motivated source geometry as in the base of a jet in AGN \citep{BlandfordZnajek1977,FalckeBiermann1995,Ghisellini2004}. The simulation of conical sources can be done utilizing the same basis vectors previously derived for a source rotating with $\Omega = d\varphi/dt$ and by constraining a sum of point sources within a region defined with height $z$ and opening angle $\psi$ (Figure {\ref{fig:conesrc}). 
	
Neither the cylindrical slab nor spheroidal geometries provide an appropriate comparison to the conical source as they are both much larger in volume. Conical sources with small opening angles $\psi$ are more plausible candidates for jet-like source bases as too great of an angle would produce a much weaker focusing of the rays in the direction of opening. Therefore, point sources will be used as the basis for comparison throughout this section. 

In Figure \ref{fig:conesrc_pt_comp} two conical geometries of height $1.5r_g \leq z \leq 5r_g$ and $5r_g \leq z \leq 10r_g$ with opening angle $\psi = 10^{\circ}$ are compared to two point sources at heights $z = 5r_g$ and $z = 10r_g$. The comparison of the emissivity profiles show that in both cases the point source located at the top of the respective cone produces a profile shape exceptionally similar to that of the matching conical source. This result is not entirely unexpected as the opening angle of the conical sources must remain small and therefore the majority of the source volume is located closer to the top of the cone than the vertex. Similar findings were obtained by \cite{Dauser2013} comparing point sources to vertically extended coronae, where it was determined that a point source at some intermediate location between the base and top of a conically shaped geometry could produce a very similar emissivity profile.

This result does prompt the need to further test the difference between point sources and conical geometries in different ways. We define the reflection fraction, $R$, as:
\begin{align}
\label{eqn:reflection}
R = \frac{N_{\text{disc}}}{N_{\text{escaped}}}
\end{align}
where $N_{\text{disc}}$ is the number of photons that land on the accretion disc, producing an observed reflection spectrum, and $N_{\text{escaped}}$ is the number of photons that escape the system entirely, forming the observed power law continuum. 

Observationally, we cannot measure the number of photons incident on the accretion disc due to various processes that take place once the photon reaches the disc that result in a different number of reflected photons. Assuming that the coronal emission is constant over time, however, allows for the accretion disc to eventually reach an equilibrium state such that the energy incident on the disc from the corona is equal to the energy output by the disc. Thus an approximation to the fraction of photons that are incident on the disc can be obtained from a reflection fraction defined as the ratio of reflected to continuum flux.

Using Equation \ref{eqn:reflection} it is found that for the conical source with height $5r_g \leq z \leq 10r_g$ (Figure \ref{fig:conesrc_pt_comp}) $R = 1.775$ and for the point source at $z = 10r_g$ that $R = 1.487$. Furthermore, for the cone at $1.5 \leq z \leq 5r_g$ $R = 3.246$ and for the point source at $z = 5r_g$ that $R = 2.114$. The reflection fraction of conical geometries are not significantly different from point sources if they are sufficiently far away from the black hole. In the case of the sources within $5r_g$ of the black hole, the extended nature of the source down to $z = 1.5r_g$ does come into effect as more photons here are focused back down onto the innermost regions of the accretion disc whereas in the case of the point source all of the photons emitted are at $z = 5r_g$ with a much reduced effect of gravity from the singularity. 

As in the previous sections we may begin the further analysis of conical sources by varying the height above the accretion disc and black hole, as seen in Figure \ref{fig:conesrc_height_comp}. The results produced here are consistent with expectations: a source closer to the black hole exhibits a nearly featureless profile with less photons landing on the outer disc while a source further away from the black hole accentuates the underlying twice-broken nature of the profile with a flattened midsection. 

Further exploration naturally leads to the variance of the opening angle of the cone, which has the effect of increasing the radial extent over the accretion disc. The results shown in Figure \ref{fig:conesrc_angle_comp} were produced by a cone at $5r_g \leq z \leq 10r_g$ and indicate that by decreasing the opening angle the photons become further focused on the innermost radii of the accretion disc while increasing the value of $\psi$ allows for more photons to land at larger radii. In fact, the conical source with the smallest angle produced an emissivity profile more similar to a point source while the profile produced by the cone with a larger opening angle resembles an extended profile such as those seen in the slab cases. Between the three examples shown in Figure \ref{fig:conesrc_angle_comp} the profile shape differs far less in comparison to the effect of height.

%########################%
\section{Beamed Sources} %
%########################%
\label{sect:beamsrc}

Beamed sources are especially relevant in AGN where outflows or jet-like structures are present. It is therefore important and interesting to study how the emissivity profile changes due to beaming and what properties of the corona can be determined from these changes. Point source and conical geometries are the most physically plausible examples of sources that would support beaming. In such coronae, the particles that comprise the source are moving in a common direction with some velocity, changing the observed properties of the system. It is the goal of these next sections to analyse the emissivity profiles produced by such coronae and examine their differences and similarities. 

%#########################%
\subsection{Point Sources} %
%#########################%
\label{sect:beampt}

\begin{figure}
	\subfloat[\label{fig:beampt}]{\includegraphics[width=\linewidth]{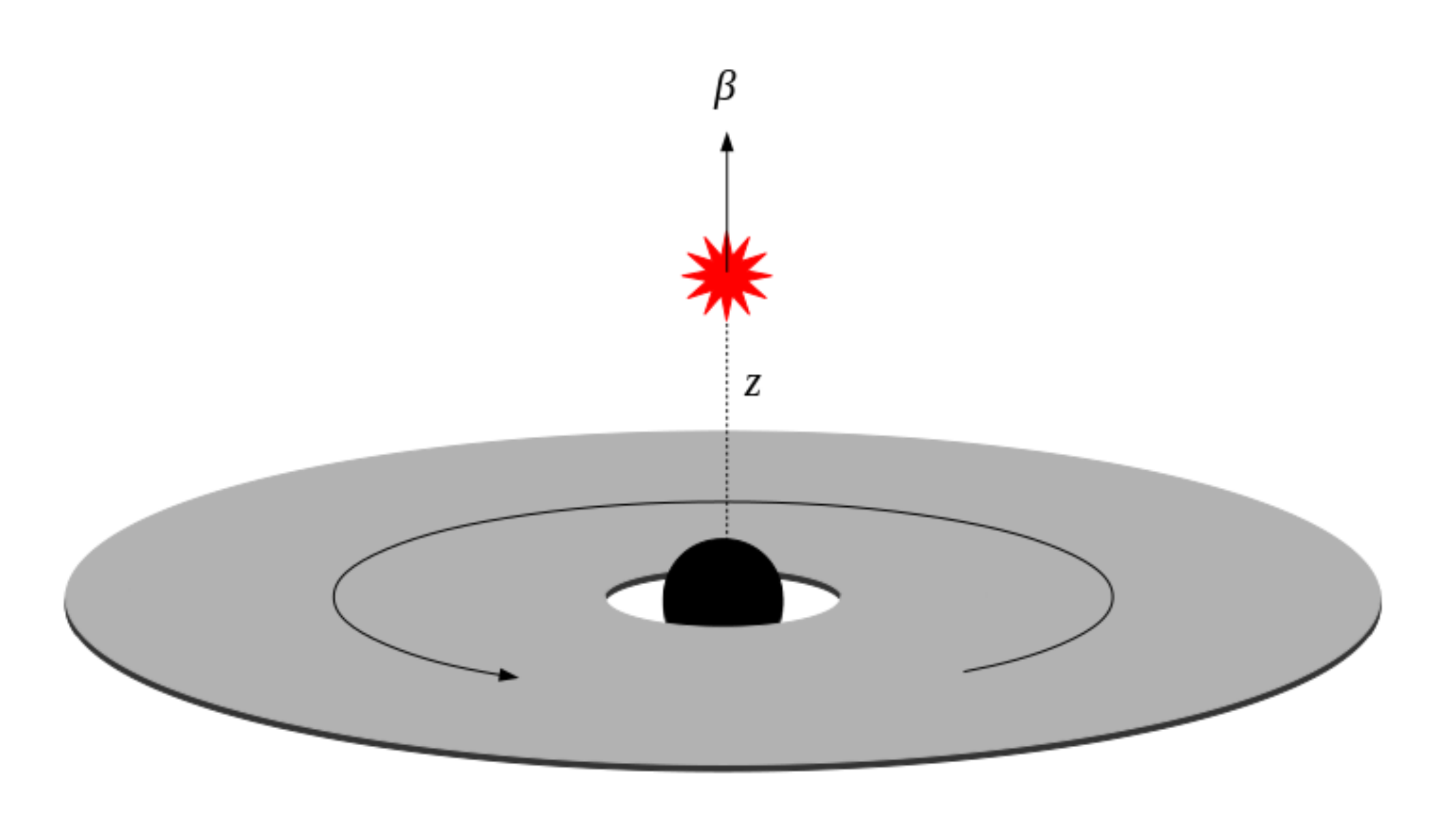}}
	\\
	\subfloat[\label{fig:beampt_emissivity_plot}]{\includegraphics[width=0.90\linewidth]{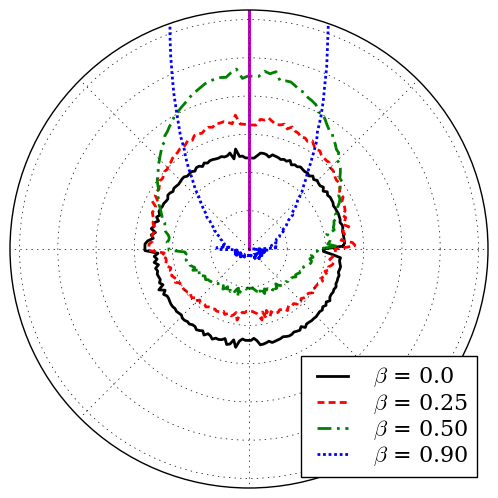}}
	\\
	\subfloat[\label{fig:beampt_beam_comp}]{\includegraphics[width=\linewidth]{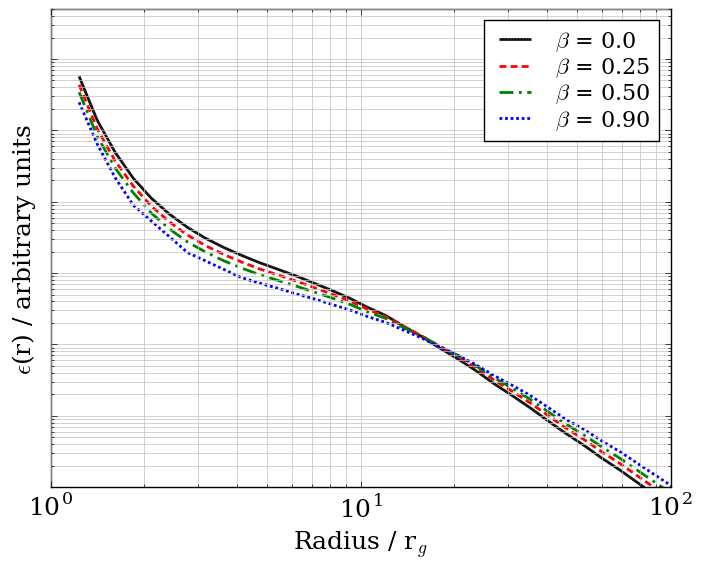}}
		
	\caption{(a) A beamed point source corona geometry with velocity $\beta$ located at a height $z$ above a rotating black hole and accretion disc. (b) A point source at height $z = 5r_g$ with various $\beta$ values showing the distribution of initial ray momenta with the magenta line representing the direction of the source velocity.  (c) A comparison of beamed point source coronae with varying $\beta$ values located a height $z = 10r_g$ above a black hole.}
	\label{fig:beampt:all}
\end{figure}

It is necessary to once again start with a simplistic point source (Figure \ref{fig:beampt}) where the corona is located at height $z$ above a black hole and is given a velocity radially away from the black hole, denoted henceforth as $\beta$. In these beamed sources, only radial motion is desired. The velocity of the source may be defined as that seen by a nearby freely falling zero angular momentum observer (ZAMO) with velocity $\omega$. We consider, in general, a ZAMO so as to be valid in the off-axis case, but we take the limit of $\omega=0$ as $\theta\rightarrow0$ (i.e. on the spin axis of the black hole) for simplicity. The observed velocity $\beta$ can therefore computed by using $ds^2 = 0$ for photons in the Kerr metric in Boyer-Lindquist co-ordinates with $d\theta = d\varphi = 0$ as (recalling that $c = \mu = 1$):
\begin{align}
\label{eqn:zamometric}
{ds}^2 &= \left(1-\frac{2r}{\rho^2}\right){dt}^2-\frac{\rho^2}{\Delta}{dr}^2 = 0
\end{align}
From here it is possible to rearrange and solve for the co-ordinate velocity $dr/dt$, which yields:
\begin{align}
\label{eqn:conversion_factor}
\left.\frac{dr}{dt}\right|_{\mathrm{photon}} &= \frac{r^2 - 2r + a^2}{r^2 + a^2}
\end{align}
This expression for $dr/dt$ provides co-ordinate velocity given the velocity $\beta$ measured in the frame of co-located ZAMO as some fraction of $c$. Therefore, the particle velocity in the source frame (i.e. source velocity measured by a co-located ZAMO) is computed as:
\begin{align}
\label{eqn:beta_src}
\left.\frac{dr}{dt}\right|_{\mathrm{src}} &= \beta \left.\frac{dr}{dt}\right|_{\mathrm{photon}}
\end{align} 
The value of $dr/dt$ in the source frame is calculated at each coordinate in the geometry, in this case at one single point source, and the observed velocity $\beta$ is given a direction radially outward from the black hole. With this understanding of beaming a new set of basis vectors for beamed sources is constructed:

\begin{align}
\begin{split}
\label{eqn:beam-basis}
\mathbf{e}'_{(t)} &= \sqrt{\frac{1}{g_{tt}+v^{2}g_{rr}}} \left( 1, v, 0, 0 \right) \\
\mathbf{e}'_{(1)} &= \sqrt{\frac{1}{g_{tt}+v^{2}g_{rr}}} \left( v\sqrt{\frac{-g_{rr}}{g_{tt}}}, \sqrt{\frac{-g_{tt}}{g_{rr}}}, 0, 0 \right) \\
\mathbf{e}'_{(2)} &= \left( 0, 0, \sqrt{\frac{-1}{g_{\theta\theta}}}, 0 \right) \\
\mathbf{e}'_{(3)} &= \sqrt{\frac{1}{g_{tt}\left(g^{2}_{\varphi t} - g_{tt} g_{\varphi\varphi}\right)}} \left( g_{\varphi t} , 0, 0, -g_{tt} \right) \\
\end{split}
\end{align}
where $\mathbf{e}'_{(t)} = \mathbf{v}$ is the four-velocity of the source moving in the $r$ direction away from the black hole and $v=\left.\frac{dr}{dt}\right|_{\mathrm{src}}$ is the radial coordinate velocity of the source.

Using the basis vectors in Equation \ref{eqn:beam-basis}, a set of simulations were run for point sources with various velocities directed away from the black hole. To ensure correctness of the beaming Figure \ref{fig:beampt_emissivity_plot} was produced showing the distribution of initial ray momenta for a source located at $z = 5r_g$, where the magenta line represents the direction of the source velocity. As the velocity increases, the number of rays with initial momenta in the source velocity direction increases, with the case where $\beta = 0.90$ produces very few rays moving down towards the accretion disc initially.

The emissivity profiles in Figure \ref{fig:beampt_beam_comp} were calculated for point sources at $z = 10r_g$ and various $\beta$ values. By increasing the velocity, fewer photons land on the inner disc while an increased number are shown to land at larger radii on the accretion disc. This result is precisely what one would expect: as $\beta$ increases, photons are preferentially given a velocity in the beaming direction, therefore acting to pull ray trajectories that were sent down toward the inner disc region out to larger radii (e.g. \citealt{Dauser2013}).

\begin{figure}
	\subfloat[\label{fig:beampt_reflection_comp}]{\includegraphics[width=\linewidth]{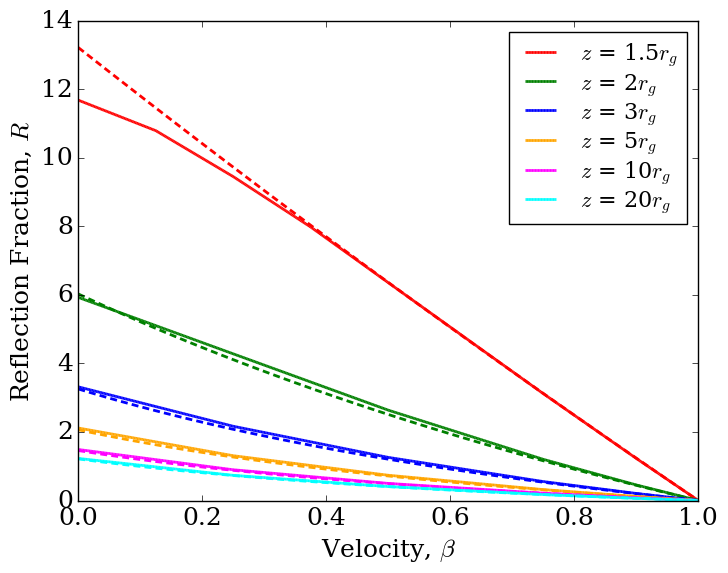}}
	\\
	\subfloat[\label{fig:beampt_reflfrac_velocity}]{\includegraphics[width=\linewidth]{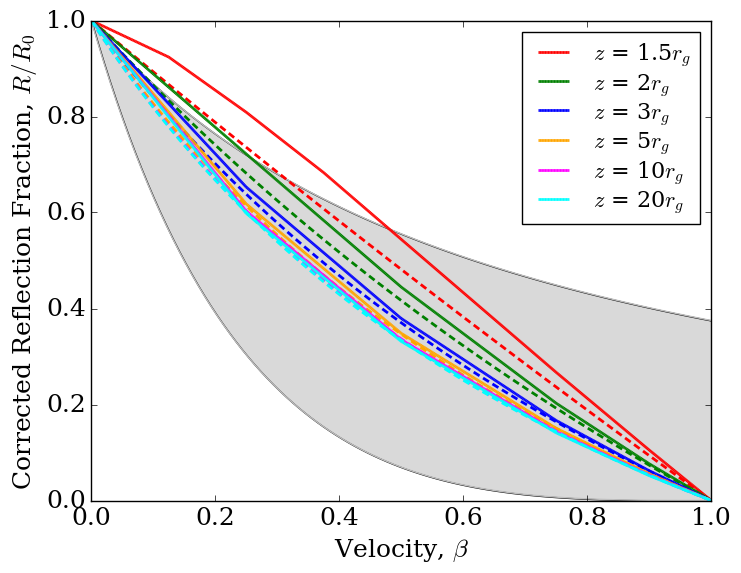}}
	\\
	\subfloat[\label{fig:beampt_reflfrac_height}]{\includegraphics[width=\linewidth]{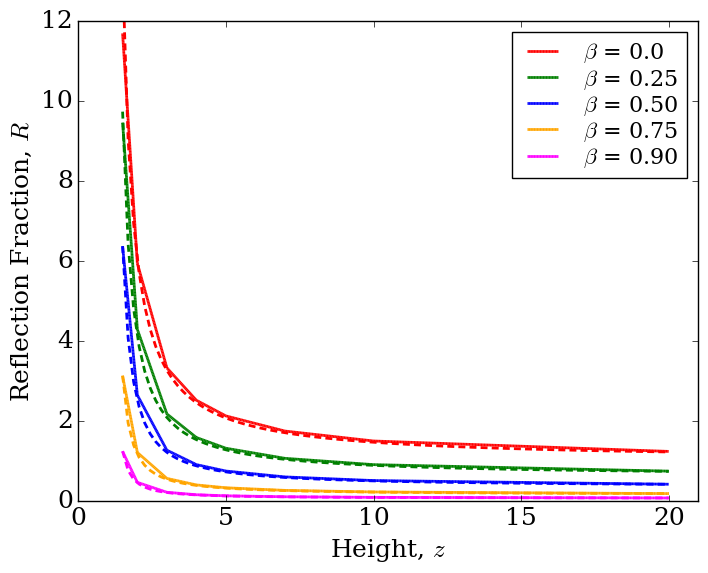}}
	
	\caption{(a) The calculated reflection fractions produced by various beamed point source coronae with varying $\beta$ values and heights ranging between $z = 1.5r_g$ and $z = 20r_g$. (b) Reflection fractions for the same set of point sources normalized by the reflection fraction at $\beta = 0$ for each source. (c) Reflection fractions for the same set of point sources with varying $\beta$ values as a function of source height. Throughout (a), (b), and (c) the dashed lines represent approximations to $R$ produced by Equation \ref{eqn:reflection_beam_full}.}
	\label{fig:beampt:refl}
\end{figure}

These differences, however, are not of the same magnitude as those seen throughout Sections \ref{sect:ptsrc} and \ref{sect:extsrc}, and thus it is necessary to examine the effects of beaming more closely using the reflection fraction. Here the reflection fraction may be defined as:
\begin{align}
\label{eqn:reflection_beam}
R = \frac{\mu_{\mathrm{out}}-\mu_{\mathrm{in}}}{1-\mu_{\mathrm{out}}}
\end{align}
where $\mu_{\mathrm{out}}$ is the cosine of the emission angle in the source frame for rays that reach the outer disc (escape the system) and $\mu_{\mathrm{in}}$ is the cosine of the emission angle for rays that reach the ISCO. This new definition of reflection fraction is equivalent to Equation \ref{eqn:reflection}. Including relativistic aberration due to the source velocity the emission angles become:
\begin{align}
\label{eqn:aberration}
\mu' = \frac{\mu - \beta}{1-\beta\mu}
\end{align}
We may approximate $\mu_{\mathrm{out}}$ as:
\begin{align}
\label{eqn:mu_out}
\mu_{\mathrm{out}} = \frac{2z}{z^2 + a^2}
\end{align}
and $\mu_{\mathrm{in}}$ as:
\begin{align}
\label{eqn:mu_in}
\mu_{\mathrm{in}} = \frac{2 - z^2}{z^2 + a^2}
\end{align}
where $z$ is the source height. Both of the above approximations are valid only for maximally spinning Kerr black holes, which are exclusively considered throughout this work. Replacing the approximations for $\mu_{\mathrm{in}}$ and $\mu_{\mathrm{out}}$ in Equation \ref{eqn:reflection_beam} and accounting for the aberration correction using Equation \ref{eqn:aberration} provides an approximation for the reflection fraction as a function of source height and velocity:
\begin{align}
\label{eqn:reflection_beam_full}
R\left(z,\beta\right) = \frac{\mu'_{\mathrm{out}}-\mu'_{\mathrm{in}}}{1-\mu'_{\mathrm{out}}}
\end{align}

Computing $R$ as in Equation \ref{eqn:reflection} for a variety of point sources at different heights and with varying $\beta$ values produces the results in Figure \ref{fig:beampt_reflection_comp}. In the case of the point source at $z = 1.5r_g$ being so close to the black hole results in most of the ray trajectories being focused on the innermost region of the disc with reflection fraction decreasing in a nearly linear fashion as the velocity is increased. For sources further away from the influence of the black hole the trend is similar as $R$ tends to zero for $\beta = 1$ though the decreasing slope is much more shallow. 

In Figure \ref{fig:beampt_reflfrac_velocity} each of the curves in Figure \ref{fig:beampt_reflection_comp} has been divided by the reflection fraction of each source at $\beta = 0$ in order correct for the beaming. Sources with $z \geq 3r_g$ exhibit corrected reflection fraction curves as a function of velocity that are very similar to each other, with those for $z \geq 5r_g$ being nearly indistinguishable. The shaded grey region corresponds to curves for the minimum and maximum inclination produced by the estimation of $R$ from \cite{Beloborodov1999} for special relativistic cases.

Figure \ref{fig:beampt_reflfrac_height} displays the relationship between reflection fraction and source height for sources with varying velocities. As the velocity increases, the impact of height on $R$ is reduced significantly.

Throughout Figure \ref{fig:beampt:refl} the dashed lines correspond to the approximation for $R$ in Equation \ref{eqn:reflection_beam_full} with colours matching the corresponding source heights (Figures \ref{fig:beampt_reflection_comp} and \ref{fig:beampt_reflfrac_velocity}) and velocities (Figure \ref{fig:beampt_reflfrac_height}). It can be seen that the approximation fits the data well in all cases where $z > 2r_g$, corresponding to the limit from \cite{Beloborodov2002}.

With the height of the source known, either from the emissivity profile or time-lag analysis and a measured value for the reflection fraction the source velocity may be calculated via rearrangement of Equation \ref{eqn:reflection_beam_full}. As aforementioned, sources closer to the black hole experience more extreme effects due to the strong gravity and therefore do not follow the given relationship as closely.

With this brief overview of beaming effects on point sources, more physical jet-like coronae may be studied via the simulation of beamed conical geometries. 

%####################%
\subsection{Conical} %
%####################%
\label{sect:beamcone}

\begin{figure}
	\subfloat[\label{fig:beamcone}]{\includegraphics[width=\linewidth]{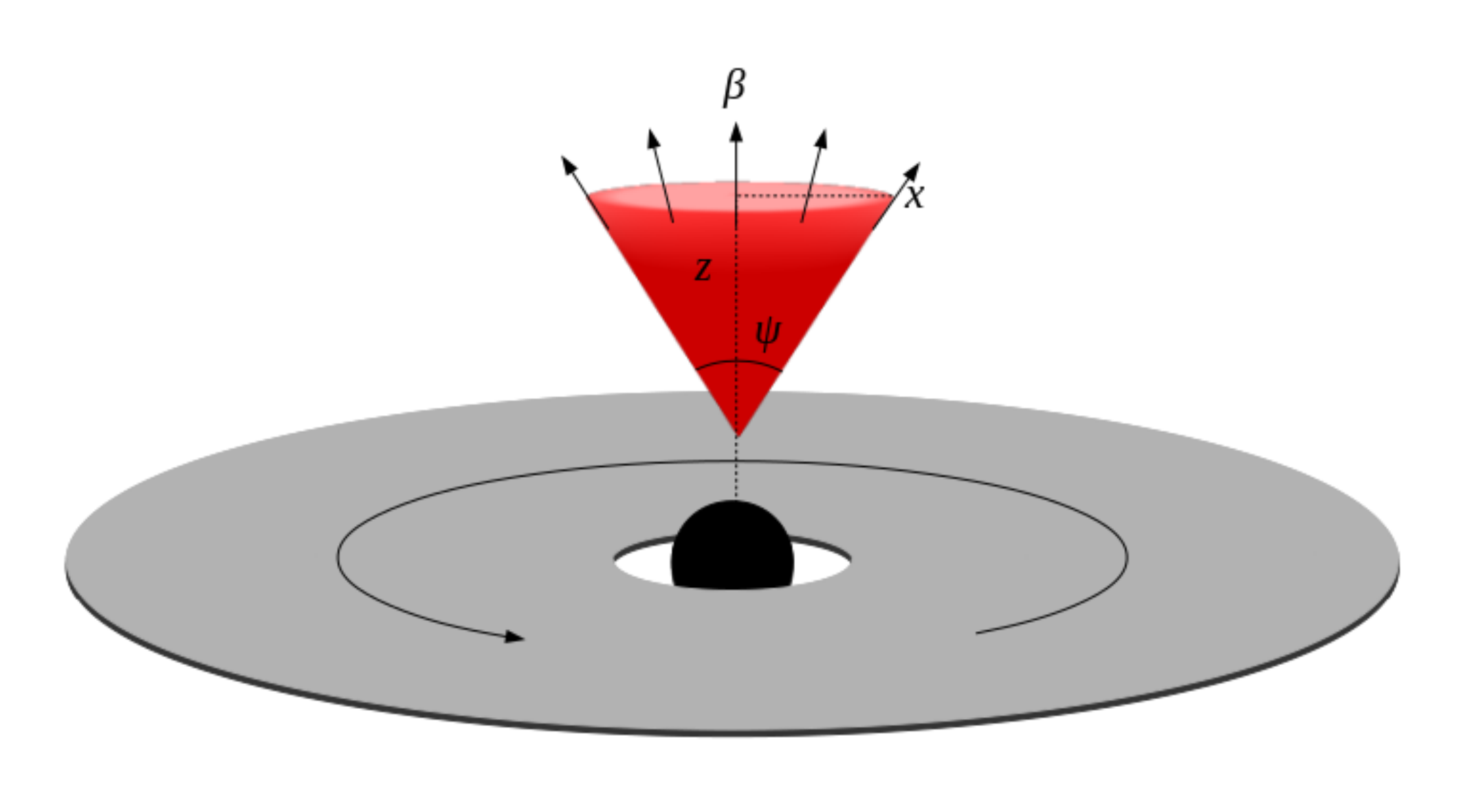}}
	\\
	\subfloat[\label{fig:beamcone_beam_comp}]{\includegraphics[width=\linewidth]{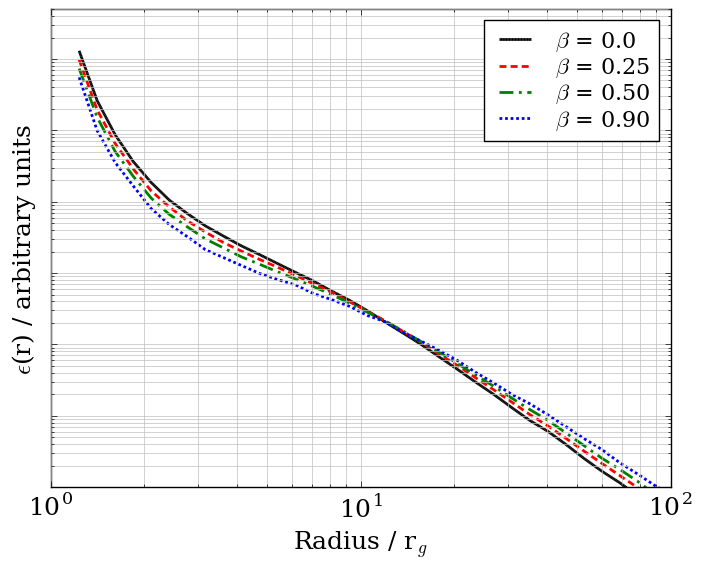}}
	\\
	\subfloat[\label{fig:beamcone_reflection_comp}]{\includegraphics[width=\linewidth]{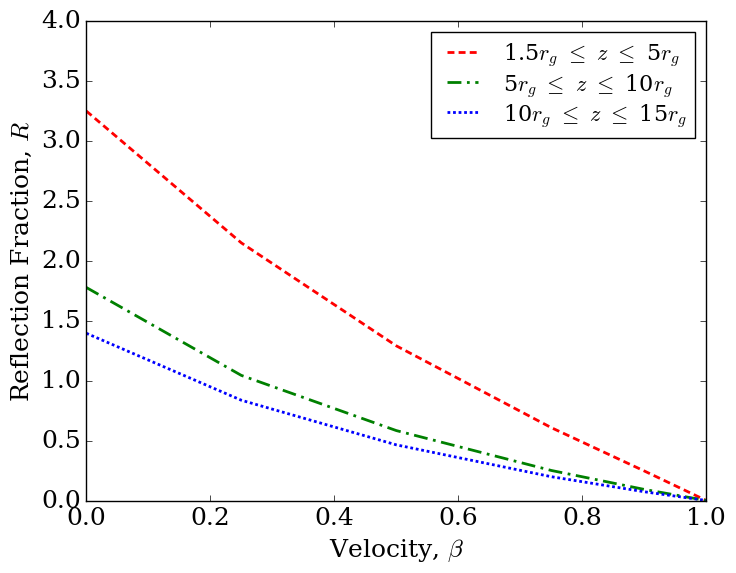}}
	
	\caption{(a) A beamed conical corona geometry located at a distance $z$ above a rotating black hole with height $z$ and opening angle $\psi$ that extends a distance $x$ over the accretion disc with velocity $\beta$. (b) A comparison of beamed conical coronae with varying $\beta$ values all with opening angle $\psi = 10^{\circ}$ located a height $5r_g \leq z \leq 10r_g$ above a black hole. (c) The calculated reflection fractions produced by beamed conical sources with varying $\beta$ values at three different heights all with opening angle $\psi = 10^{\circ}$.}
	\label{fig:beamcone:all}
\end{figure}

A conical source beamed radially away from the black hole with a small opening angle is physically plausible as the jet in AGN. Figure \ref{fig:beamcone} shows a schematic of the simulations run for beamed conical sources with a cone of height $z$ with opening angle $\psi$ and velocity $\beta$. It is important to note that the rest frame velocity is a constant throughout the extended geometry according to an observer. That is to say, points in the conical geometry that are nearest the black hole will have a source velocity greater than those furthest away, though due to how Equation \ref{eqn:conversion_factor} changes based on distance from the black hole the value of $\beta$ that is observed will remain constant through the region. 

Taking a conical source with $5r_g \leq z \leq 10r_g$ and $\psi = 10^{\circ}$ and simulating for various values of $\beta$ produces the emissivity profiles shown in Figure \ref{fig:beamcone_beam_comp}. These results are consistent with those obtained in the previous section for beamed point sources. By increasing $\beta$ fewer photons land on the inner portion of the accretion disc and a greater number reach larger radii. The mechanism here is also the same: ray trajectories that were originally headed down toward the accretion disc in the un-beamed case are pulled up and away for larger values of $\beta$. 

Performing a similar reflection fraction analysis as with the beamed point sources produces the result shown in Figure \ref{fig:beamcone_reflection_comp}. The trend exhibited is as expected from the previous analysis with a smoothly decreasing reflection fraction for increased value of $\beta$. Similarly to the beamed sources, close proximity to the black hole results in drop-offs in $R$ for higher velocities while moving further away from the gravitational influence of the singularity produces smoother curves that decrease in a near linear fashion. Comparing the results in Figures \ref{fig:beampt_reflection_comp} and \ref{fig:beamcone_reflection_comp} for point sources located at the top of each of the cones simulated reveals that the reflection fraction for conical geometries drops off more quickly than in the case of point sources. This is due to the extended nature of the cone in which the portion of the source located furthest away from the black hole emits more photons capable of escaping, lowering the reflection fraction.

%##############################%
\section{Modelling with XSPEC} %
%##############################%
\label{sect:XSPEC}

In this section we revisit all of the corona geometries that have been covered throughout this paper in an attempt to distinguish each from the other by making use of the resulting emissivity profiles in spectral modelling. This was made  possible by using \textsc{xspec} \citep{Arnaud1996} to produce a model composed of a power law component and a reflection spectrum using \textsc{reflionx} \citep{Ross2005} convolved with \textsc{kdblur} to blur the spectrum in the 0.1--100 keV range. Here \textsc{kdblur} makes use of the computed values of the emissivity profiles produced through the ray tracing simulations. In addition, each of the two components were set to equal flux through the use of \textsc{cflux}. 

For all of the models presented in this section the parameter values listed in Table \ref{tab:model_params} were used and held constant across each, only allowing for the values of the emissivity profile to change in the \textsc{kdblur} component for the various source geometries. The range was truncated between 0.3--70 keV to show the regions of interest. Parameter values in Table \ref{tab:model_params} were selected to simulate a NLS1 type galaxy (like 1H~0707) with a steep photon index, moderately high iron abundance, and a strong Fe K$\alpha$ line. 
\begin{table}
	\begin{center}
		\caption{Model components and their respective parameter values used to produce the spectra throughout Section \ref{sect:XSPEC}.}
		\begin{tabular}{ccccccc}                
			\hline
			Model Component & Parameter & Value \\
			\hline
			\hline
			powerlaw & Photon index, $\Gamma$ & 2.5 \\
			\hline
			kdblur & Inclination, $i$ & $60^{\circ}$ \\
			\hline
			reflionx & Photon index, $\Gamma$ & 2.5 \\
			& Iron abundance / solar & 5 \\
			& Ionization parameter, $\xi$ & $50 \, \mathrm{erg \, cm / s}$ \\
			\hline
			cflux & Reflection fraction, $R$ & 1.0 \\
			\label{tab:model_params}
		\end{tabular}
	\end{center}
\end{table}
It is important to note that only $i = 60^{\circ}$ is included as this inclination provides the most variance among the spectra, allowing for a best-case scenario in being able to distinguish different source geometries.

\begin{figure}
	\scalebox{1.0}{	\includegraphics[width=\linewidth]{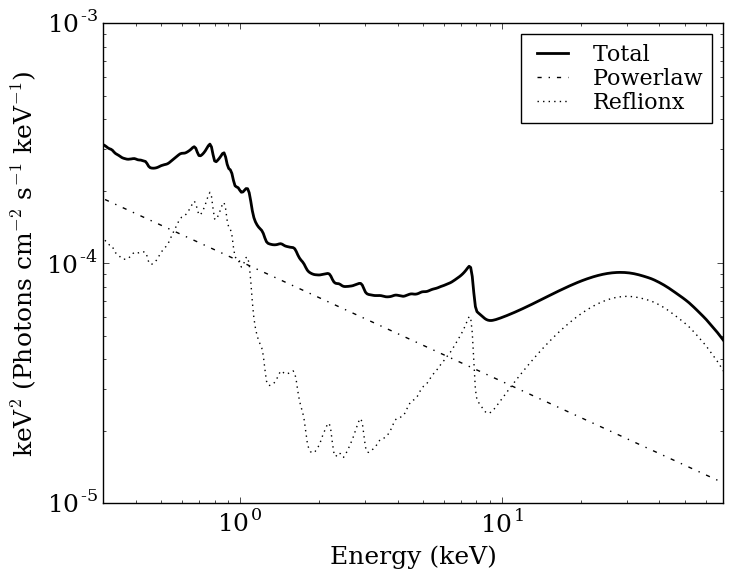}}
	\caption{Resulting spectrum produced by the model in Table \ref{tab:model_params} using the emissivity profile of a point source corona at $z = 5r_g$. The reflection and power law components are made visible for completeness.}
	\label{fig:xspec_pt5rg_allcomps}
\end{figure}

\begin{figure}
	\subfloat[\label{fig:xspec_sphcyl}]{\includegraphics[width=\linewidth]{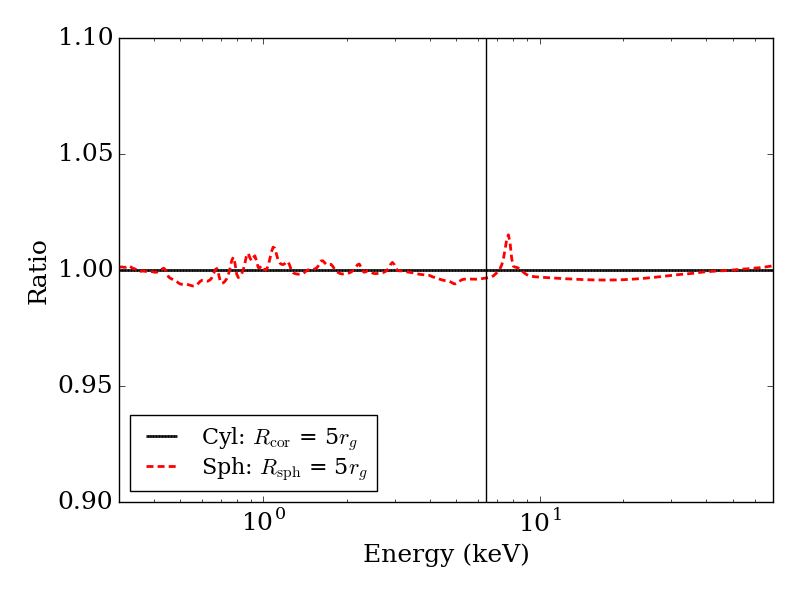}}
	\\
	\subfloat[\label{fig:xspec_conept}]{\includegraphics[width=\linewidth]{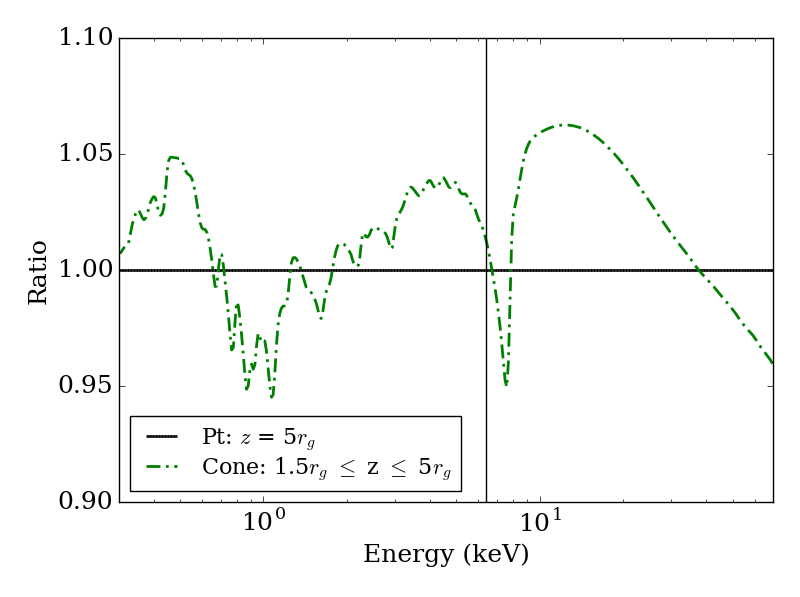}}
	\\
	\subfloat[\label{fig:xspec_allz5}]{\includegraphics[width=\linewidth]{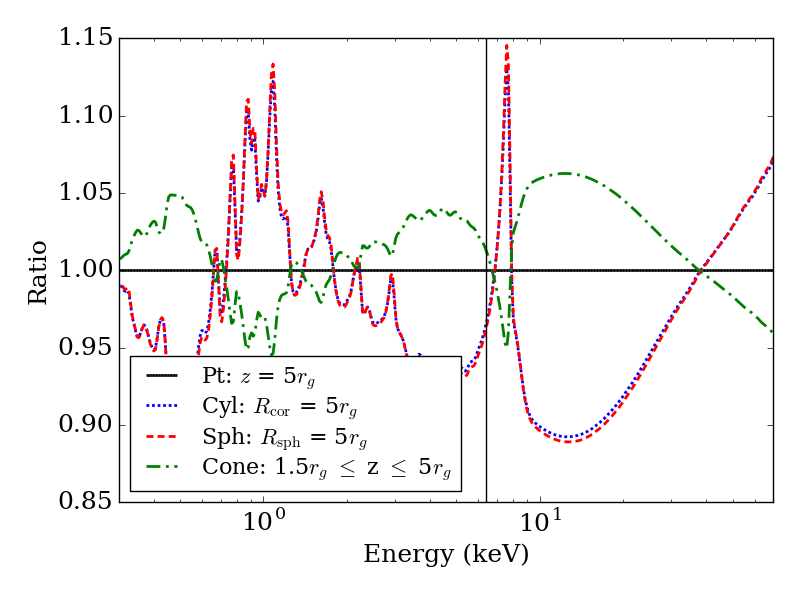}}
	
	\caption{(a) The black line denotes the ratio of the spectrum produced by a cylindrical source with itself and the red line denotes the ratio of a similar sized spheroidal source with this cylindrical geometry. (b) The ratios between a point source and itself as well as a conical geometry and the same point source. (c) The ratios of all sources within $5r_g$ of the black hole with a point source located at $z = 5r_g$.}
	\label{fig:xspec:ratios}
\end{figure}

\begin{figure}
	\subfloat[\label{fig:xspec_beamptz5}]{\includegraphics[width=\linewidth]{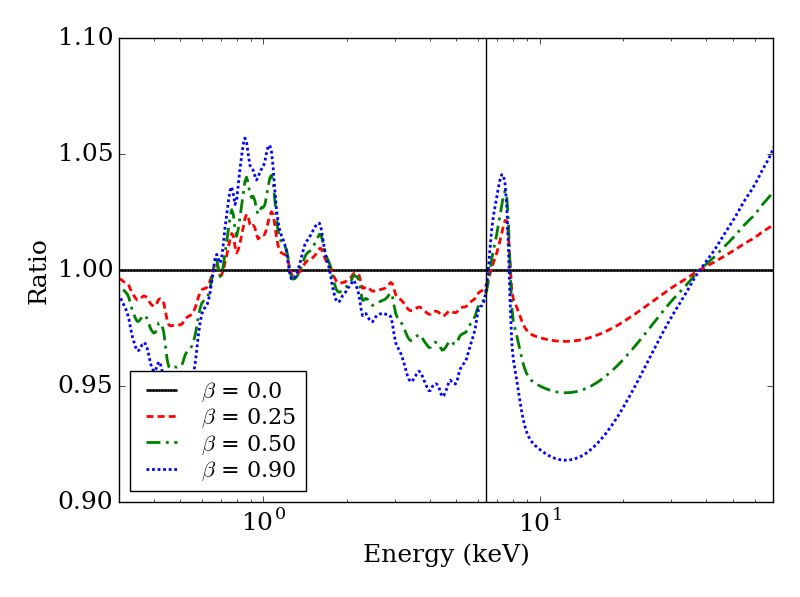}}
	\\
	\subfloat[\label{fig:xspec_beamconez1p5-5}]{\includegraphics[width=\linewidth]{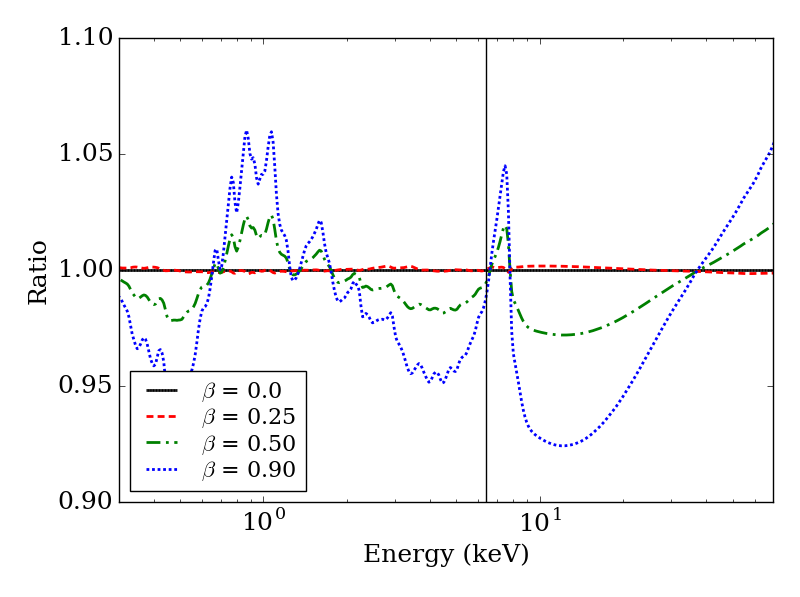}}
	
	\caption{(a) The ratios between a point source at $z = 5r_g$ and point sources at the same height with varying values of $\beta$. (b) A conical source at $1.5r_g \leq z \leq 5r_g$ with the ratios taken between conical sources at the same height with varying values of $\beta$.}
	\label{fig:xspec:beamratios}
\end{figure}

The first step in gaining an understanding as to how the spectra change for various geometries is to examine the preliminary results of this spectral modelling for a simple point source at $z = 5r_g$ (Figure \ref{fig:xspec_pt5rg_allcomps}). At 6.4 keV the prominent Fe K$\alpha$ line is present and broadened, as are the other lines in the spectrum, by the \textsc{kdblur} model. The broadened line exhibits a peaked blue wing and an extended red wing.

With this basic picture for our most simple model the other more complex extended geometries may now be explored. The variations between the spectra of the different sources, however, are not easily visible through the spectra alone. Therefore, for each case a ratio has been taken between two sources of similar geometry and/or size (slabs are comparable to spheroidal sources and points to cones). 

These ratio plots are presented in Figures \ref{fig:xspec_sphcyl} and \ref{fig:xspec_conept} where the vertical lines represent the Fe K$\alpha$ line for reference (this is included in all ratio plots). In Figure \ref{fig:xspec_sphcyl} we can see that the ratio between the spheroidal corona with $R_{\mathrm{sph}} = 5r_g$ and a like sized slab geometry remains within a few per cent throughout the 0.3--70 keV range of the plot. The most notable differences are present in the blue wing of the Fe K$\alpha$ line with the spheroidal source showing a peak at energies just beyond 6.4 keV in the 7--8 keV range, though even here the differences are small. 

The plot shown in Figure \ref{fig:xspec_conept} with the ratio taken between a conical source at $1.5r_g \leq z \leq 5r_g$ and a point source at $z = 5r_g$ shows that the spectrum produced by a conical geometry has larger ratios in the red wing of the Fe K$\alpha$ line and smaller ratios in the blue wing compared to a point source at similar height, with differences over the two spectra within 5 per cent.

Figure \ref{fig:xspec_allz5} presents all sources within $5r_g$ of the black hole and takes the ratio of these spectra with a point source at $z = 5r_g$. It can be seen that the point source is more similar to the conical geometry with the ratios of the slab and spheroidal geometries with the point source in the range of 10--15 per cent difference maximum.

\begin{figure}
	\scalebox{1.0}{	\includegraphics[width=\linewidth]{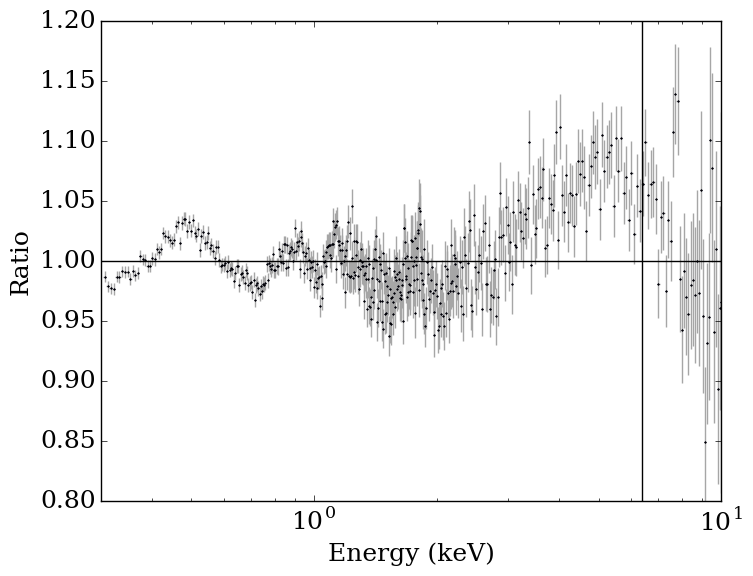}}
	\caption{The ratio of a spectrum produced by a point source corona at $z = 5r_g$ to the model of a spherical geometry of radius $R_\mathrm{sph} = 5r_g$ located on the accretion disc that has been fit to the simulated data.}
	\label{fig:xspec_sphfit2point_5rg}
\end{figure}

Both Figures \ref{fig:xspec_beamptz5} and \ref{fig:xspec_beamconez1p5-5} show similar results. In the case of beamed sources versus stationary sources, the blue wing of the Fe K$\alpha$ has been accentuated as a result of the beaming with a large peak in the spectrum ratios between 7--8 keV. Only the cases with the highest value of $\beta$ show differences in the ratios in the range of 5--10 per cent. Simulations of spectral data, as it would be measured by \xmm, were performed based on extreme objects with properties given in Table \ref{tab:model_params} that are bright (2--10 keV flux $\sim 10^{-11}$ erg/cm$^2$/s) and observed over long exposure times (e.g. 100 kiloseconds) to produce good quality spectra. For such sources the geometries are distinguishable from each other, as shown by the ratio plot in Figure \ref{fig:xspec_sphfit2point_5rg}. This solution may not be unique from other interpretations of the AGN spectrum (e.g. partial covering or multiple Comptonising layers), but it suggests that current data could distinguish between different geometries.

%####################%
\section{Discussion} %
%####################%

The emissivity profiles obtained via ray tracing simulations as well as the spectral modelling of point sources and extended corona geometries provide insight into the differences and similarities between the four geometries studied. In all of the emissivity profiles calculated a twice-broken power law shape is observed with steep slope at low radii, flattened midsection, and outer profile slope dropping off as $r^{-3}$ . After simulating and confirming the results of simple ``lamppost'' point source models a series of three extended geometries were studied by varying a number of parameters.

The height of the source above the black hole was found to cause the greatest difference in profile shape for point source and conical geometries (Figures \ref{fig:ptsrc_height_comp} and \ref{fig:conesrc_height_comp}) while being less significant the radially extended geometries (Figures \ref{fig:cylsrc_height_comp}, \ref{fig:sphsrc_cyl_comp}). In all cases increased height caused the midsection of the profile to flatten further and accentuate the twice-broken shape of the profile described in WF12. 

Horizontal displacement from the spin axis for point sources (Figure \ref{fig:ptsrc_disp_comp}) and radial extent over the accretion disc for conical geometries (Figure \ref{fig:conesrc_angle_comp}) also produced notable profile differences though the results here were much less distinct than for vertical height. In the case of cylindrical and spheroidal sources radial extent had a greater effect on the profile shape (Figures \ref{fig:cylsrc_extent_comp} and \ref{fig:sphsrc_cyl_comp}) being a more significant parameter than the height of such sources. For geometries with more total coverage over the accretion disc the break in the emissivity profiles shifted to larger radii, effectively tracing the edge of the extended source.

While being able to tell what properties a particular geometry possesses it is perhaps more interesting to study the differences between the various sources simulated. Comparing the emissivity profiles produced by the four different geometries found that point sources can be distinguished from extended geometries such as cylindrical slabs and spheroidal clouds. 

Cylindrical and spheroidal geometries were found to produce very similar emissivity profiles with minute differences (Figure \ref{fig:sphsrc_cyl_comp}). Comparisons between spheroidal and ellipsoidal sources found no significant differences between the two to allow for any true differentiation. 

The same can be said about the differences between point source and conical geometries with narrow opening angles. In the case of conical sources with narrow opening angles the emissivity profile alone is insufficient in determining whether the parent source geometry is a point or cone (Figure \ref{fig:conesrc_pt_comp}). These results suggest that radial extent and coverage over the accretion disc are more impactful on the resulting emissivity profile than vertical extent perpendicular to the disc. 

However, differences between point source and conical geometries arise once reflection fraction is considered, with the conical source exhibiting a larger reflection fraction in all cases than the equivalent point source. This is as expected due to the extended nature of the cone, allowing more photons to be emitted nearer the black hole than a point source located at the top of the cone, resulting in fewer photons being able to escape the system and more being reflected off of the accretion disc. 

In the analysis of beamed point source and conical geometries, which were given a velocity $\beta$ directed radially away from the black hole, the emissivity profiles in both cases exhibit a common trend: flattening of the profile midsection with increased $\beta$ (Figures \ref{fig:beampt_beam_comp} and \ref{fig:beamcone_beam_comp}). The profiles for the various $\beta$ values are all very similar to each other, and thus it is again necessary to explore the reflection fraction. Changes in $R$ for point source and conical geometries follow an expected trend of decreasing to zero with a shallow slope as $\beta$ approaches the limit of light speed, with the exception being the point source located closest to the black hole (Figure \ref{fig:beampt_reflection_comp} and \ref{fig:beamcone_reflection_comp}). An approximation of the measured reflection fraction in Equation \ref{eqn:reflection_beam_full} was found to fit the data for beamed point sources with $z > 2r_g$ well. With the height of the point source known, from an analysis of the emissivity profile or through time-lag studies, and the reflection fraction measured from AGN data it is possible to rearrange Equation \ref{eqn:reflection_beam_full} and estimate the source velocity.

In Section \ref{sect:intro} it was discussed that the illumination profile (photons incident on the disc) and emissivity profile (photons processed and re-emitted by the disc) were to be used interchangeably throughout this work. The equivalency assumed, however, may not be exactly true due to various factors including the absorption and emission processes in the disc that are affected by ionisation gradient across the surface of the accretion disc which contribute to differences between the illumination and emissivity profile. For example, \cite{Svoboda2012} show the effects of ionisation gradient on the emissivity profile. In this work we were interested in emissivity profiles as measured from the Fe K$\alpha$ emission line and assumed that the line flux is proportional to the incident flux upon the disc with no other factor (e.g. ionisation gradient) causing variation in the emitted line flux between different radii.

Using the computed emissivity profiles it is possible to simulate spectra using \textsc{xspec} to explore the feasibility of distinguishing the different geometries in observational AGN spectral data. Modelling in \textsc{xspec} was done using a combination of power law and blurred reflection models. This model was used to create a spectrum for each corona geometry that could then be compared with the spectrum of another geometry in an effort to examine differences between them (Figure \ref{fig:xspec:ratios}). 

With this technique it can be seen that the largest differences between the spectra occurs around the Fe K$\alpha$ line at 6.4 keV. Figure \ref{fig:xspec_allz5} shows that in a comparison of all geometries to a point source the differences are less than 15 per cent, with cylindrical and spheroidal sources being most different. Performing the same analysis with beamed point sources and conical geometries finds that differences between the spectra are about 5--10 per cent when comparing a stationary source to beamed sources of the same geometry. 

The results found throughout Section \ref{sect:XSPEC} suggest moderate differences in the spectra for the various geometries studied. The differences require high quality data from current missions observing extreme, bright sources over long exposure times to be distinguishable. The solution found here may not be unique as other model interpretations could result in good fits. Athena-like missions with large collecting area (e.g. \citealt{Nandra2013}) or those that are Hitomi-like with high spectral resolution (e.g. \citealt{Takahashi2014}) will provide the ability to more finely discern the differences between the geometries.

All of the results obtained throughout this study of various corona geometries suggest that emissivity profiles are a useful tool in determining different sources, but that they are also part of a larger tool-kit. The corona geometry itself is an important indicator of the formation processes that take place to create the X-ray source in AGN, processes which are still not fully understood. Parameters such as the extent of the source over the accretion disc are readily identified through the analysis of the emissivity profile. In order to precisely identify the geometry of a specific source emissivity profiles must be used in conjunction with studies in reflection fraction, time lag analysis, and spectral modelling. Beaming, for example, proved to have a lesser effect on the emissivity profile than did extent over the disc, requiring an analysis of reflection fraction in order to determine the effect of source velocity.

%#####################%
\section{Conclusions} %
%#####################%

It was found that point sources can be distinguished from extended geometries such as cylindrical slabs and spheroidal clouds through the differences in the emissivity profiles alone. 

Conical geometries were found to produce results very similar to point sources, and thus required the analysis of reflection fraction in order to distinguish the two geometries. 

In the case of beamed coronae it was again necessary to use reflection fraction in order to determine the differences between beamed geometries as the emissivity profiles were not significantly different. 

By analysing reflection fraction curves as a function of source velocity and height it was possible to produce an approximation of the point source reflection fraction as well as the velocity given the height of the corona and its measured reflection fraction.

Differences in the simulated spectra produced by different geometries do not exceed 15 per cent even in the most extreme cases. 

Simulated data suggest that differences between the emissivity profiles of the various geometries can be detected in high quality data for extreme, bright sources over long exposures.

The results all suggest that a collective tool-kit involving several analysis techniques (e.g. emissivity profiles, reflection fraction, time lag analysis, and spectral modelling) are required in order to determine the true geometry of the corona.

%###########################%
\section*{Acknowledgements} %
%###########################%

DRW is supported by NASA through Einstein Postdoctoral Fellowship grant number PF6-170160, awarded by the Chandra X-ray Centre, operated by the Smithsonian Astrophysical Observatory for NASA under contract NAS8-03060. We thank the anonymous referee for their careful reading and helpful comments on the original manuscript. AGG would like to thank D. F. Casta{\~n}eda for assistance with preparing figures.

%%%%%%%%%%%%%%%%%%%%%%%%%%%%%%%%%%%%%%%%%%%%%%%%%%


\bibliographystyle{mnras}
\bibliography{refs} % if your bibtex file is called example.bib

% Don't change these lines
\bsp	% typesetting comment
\label{lastpage}
\end{document}